\documentclass[sn-nature]{sn-jnl}

\usepackage{graphicx}%
\usepackage{multirow}%
\usepackage{amsmath,amssymb,amsfonts}%
\usepackage{amsthm}%
\usepackage{mathrsfs}%
\usepackage[title]{appendix}%
\usepackage{xcolor}%
\usepackage{textcomp}%
\usepackage{manyfoot}%
\usepackage{booktabs}%
\usepackage{algorithm}%
\usepackage{algorithmicx}%
\usepackage{algpseudocode}%
\usepackage{listings}%

\theoremstyle{thmstyleone}%
\theoremstyle{thmstyletwo}%

\theoremstyle{thmstylethree}%

\newcommand{\grb}{GRB\,221009A}

\begin{document}

\title[Article Title]{JWST Observations of the Extraordinary GRB\,221009A Reveal an Ordinary Supernova Without Signs of r-Process Enrichment in a Low-Metallicity Galaxy}


\author*[1]{Peter K.~Blanchard}\email{peter.blanchard@northwestern.edu}
\author[2]{V.~Ashley Villar}
\author[3]{Ryan Chornock}
\author[4,5]{Tanmoy Laskar}
\author[6,7]{Yijia Li}
\author[6,7,8]{Joel Leja}
\author[9]{Justin Pierel}
\author[2]{Edo Berger}
\author[3,10]{Raffaella Margutti}
\author[11]{Kate D.~Alexander}
\author[12]{Jennifer Barnes}
\author[2]{Yvette Cendes}
\author[1]{Tarraneh Eftekhari}
\author[10]{Daniel Kasen}
\author[3]{Natalie LeBaron}
\author[13,14]{Brian D.~Metzger}
\author[9]{James Muzerolle Page}
\author[9]{Armin Rest}
\author[1,15]{Huei Sears}
\author[16,17]{Daniel M.~Siegel}
\author[2]{S.~Karthik Yadavalli}

\affil[1]{Center for Interdisciplinary Exploration and Research in Astrophysics (CIERA), Northwestern University, 1800 Sherman Ave.~8th Floor, Evanston, IL 60201, USA}

\affil[2]{Center for Astrophysics \(|\) Harvard \& Smithsonian, 60 Garden St.~Cambridge, MA 02138, USA}

\affil[3]{Department of Astronomy, University of California, Berkeley, CA 94720-3411, USA}

\affil[4]{Department of Physics \& Astronomy, University of Utah, Salt Lake City, UT 84112, USA}

\affil[5]{Department of Astrophysics/IMAPP, Radboud University, PO Box 9010, 6500 GL, The Netherlands}

\affil[6]{Department of Astronomy \& Astrophysics, The Pennsylvania State University, University Park, PA 16802, USA}
\affil[7]{Institute for Gravitation and the Cosmos, The Pennsylvania State University, University Park, PA 16802, USA}
\affil[8]{Institute for Computational \& Data Sciences, The Pennsylvania State University, University Park, PA 16802, USA}

\affil[9]{Space Telescope Science Institute, 3700 San Martin Drive, Baltimore, MD 21218-2410, USA}

\affil[10]{Department of Physics, University of California, 366 Physics North MC 7300, Berkeley, CA 94720, USA}

\affil[11]{Department of Astronomy/Steward Observatory, 933 North Cherry Avenue, Rm. N204, Tucson, AZ 85721-0065, USA}

\affil[12]{Kavli Institute for Theoretical Physics, Kohn Hall, University of California, Santa Barbara, CA 93106, USA}

\affil[13]{Department of Physics and Columbia Astrophysics Laboratory, Columbia University, Pupin Hall, New York, NY 10027, USA}

\affil[14]{Center for Computational Astrophysics, Flatiron Institute, New York, NY 10010, USA}

\affil[15]{Department of Physics and Astronomy, Northwestern University, 2145 Sheridan Road, Evanston, IL 60208-3112, USA}

\affil[16]{Institute of Physics, University of Greifswald, D-17489 Greifswald, Germany}

\affil[17]{Department of Physics, University of Guelph, Guelph, Ontario N1G 2W1, Canada}


\abstract{
\unboldmath
Identifying the astrophysical sites of the $r$-process, one of the primary mechanisms by which heavy elements are formed, is a key goal of modern astrophysics.  The discovery of the brightest gamma-ray burst of all time, \grb, at a relatively nearby redshift, presented the first opportunity to spectroscopically test the idea that $r$-process elements are produced following the collapse of rapidly rotating massive stars.  Here we present spectroscopic and photometric \textit{James Webb Space Telescope} (JWST) observations of \grb\ obtained $+168$ and $+170$ rest-frame days after the initial gamma-ray trigger, and demonstrate they are well-described by a supernova (SN) and power-law afterglow, with no evidence for an additional component from $r$-process emission, and that the SN component strongly resembles the near-infrared spectra of previous SNe, including SN\,1998bw. We further find that the SN associated with \grb\ is slightly fainter than the expected brightness of SN\,1998bw at this phase, concluding that the SN is therefore not an unusual GRB-SN.  We infer a nickel mass of $\approx0.09$ M$_{\odot}$, consistent with the lack of an obvious SN detection in the early-time data.  We find that the host galaxy of \grb\ has a very low metallicity of $\approx0.12$ Z$_{\odot}$ and our resolved host spectrum shows that \grb\ occurred in a unique environment in its host characterized by strong H$_2$ emission lines consistent with recent star formation, which may hint at environmental factors being responsible for its extreme energetics.
}

\maketitle

\section{Introduction}\label{sec1}



The origin of the heaviest elements in the Universe, specifically those formed via rapid neutron capture ($r$-process) nucleosynthesis, remains a major open question in astrophysics \cite{cowan2021origin, siegel2022r}.  Given the high density of neutron-rich material needed for the $r$-process to occur, the collisions of neutron stars have long been a suspected source \citep{1982ApL....22..143S, Eichler1989}, and indeed, the observations of the kilonova associated with GW\,170817 confirmed that binary neutron star (BNS) mergers are the source of at least some of the $r$-process material in the Universe \citep{arcavi2017optical, Cowperthwaite2017,Chornock2017, drout2017light, smartt2017kilonova}.  However, there is growing evidence that there may be multiple sites of $r$-process nucleosynthesis from studies of low-metallicity galactic halo stars, dwarf galaxy and globular cluster enrichment \citep{cote2017advanced,naiman2018first, cote2019neutron,hotokezaka2018neutron}.

A second proposed site of the $r$-process is in rapidly rotating cores of massive stars that collapse into an accreting black hole, producing similar conditions as the aftermath of a BNS merger \citep{siegel2019}.  Theoretical simulations suggest that accretion disk outflows in these so-called ``collapsars" may reach the neutron-rich state required for the $r$-process to occur \cite{siegel2019, just2022r}.  The larger mass of $r$-process material synthesized per event compared to BNS mergers suggests that collapsars could be a dominant source, making them a possible missing piece in our understanding of $r$-process enrichment in the Universe.

The discovery of the long-duration gamma-ray burst \grb, the brightest GRB ever observed \citep{burns2023grb,Lesage2023,Williams2023}, on 9 October 2022 at a relatively nearby redshift of $z=0.151$ \cite{Malesani2023} presents a unique opportunity to search for $r$-process signatures in a collapsar.  Collapsars are the favored explanation for long GRBs (LGRBs), which result from the launch of a relativistic jet and its subsequent interaction with the surrounding medium. $r$-process nucleosynthesis is more likely to occur in collapsars with large accretion disk masses, which are also thought to be linked with brighter GRBs \cite{siegel2019}, making \grb\ a particularly strong candidate to search for $r$-process signatures. These events are known to be accompanied by broad-lined Type Ic supernovae (SNe Ic-BL) characterized by higher velocities than normal Type Ic SNe, suggesting the energy powering LGRBs also impacts the associated SNe (see \cite{cano2017observer} for a review).  

It is the SN following a LGRB that would be responsible for carrying $r$-process material from the explosion site into the interstellar medium.  While early-time observations of \grb\ provided an exquisite view of the afterglow \cite{Laskar2023,OConnor2023}, to date there are conflicting claims in the literature regarding the presence of an associated supernova due in part to the bright afterglow and high Galactic extinction \cite{Shrestha2023,Fulton2023,Srinivasaragavan2023,levanboat}.  Moreover, there have been claims that two recent LGRBs are associated with BNS mergers \cite{Rastinejad2022, levan2023jwst}, making the search for a SN associated with \grb\ crucial not only for an $r$-process search, but also for understanding the origin of its extreme luminosity.    

Here, we present late-time \textit{James Webb Space Telescope} (JWST) observations of \grb\ consisting of a near-infrared (NIR) spectrum and imaging in four NIR bands.  These observations provide the first clear detection of a SN associated with this extreme event and enable the first search for $r$-process emission in a nebular phase spectrum of a GRB-SN.  Moreover, these data provide a detailed NIR view of the host galaxy, enabling an assessment of environmental factors that may be responsible for this extraordinary GRB.

\section{Identification of Supernova Emission}

We obtained spectroscopy with the Near Infrared Spectrograph (NIRSpec) using the medium-resolution gratings covering $1-3\mu$m on 20 April 2023 and imaging with the Near Infrared Camera (NIRCam) using the F115W, F200W, F277W, and F444W filters on 22 April 2023.  These observations occurred $+194$ and $+196$ observer-frame days after the burst, respectively (rest-frame phases of $+168$ and $+170$ days). The afterglow of \grb\ is clearly detected in our images, from which we measure photometry (Figure \ref{fig:images}; see Methods for details).  In our NIRSpec observations, we detect a clear spectral trace containing flux from \grb\ and its host galaxy (Extended Data Figures \ref{fig:G140M} and \ref{fig:G235M}; see Methods for details of the spectral extraction).  Due to the high Galactic extinction \cite{schlafly2011measuring}, and possible non-negligible extinction intrinsic to the host galaxy \cite{Malesani2023}, we re-analyze archival early-phase NIRSpec/PRISM and MIRI spectra \cite{levanboat} of \grb\ using multiple dust laws to constrain the extinction (see Methods and Extended Data Figure \ref{fig:prism}).  We use the resulting extinction parameters (Extended Data Table \ref{tab:dust}) to correct our rest-frame $+168$ day NIRSpec grating spectrum.  

In Figure \ref{fig:specphot} we show two versions of the spectrum, one corrected using an extinction curve from \cite{Fitzpatrick99} and another using one from \cite{gordon2023one}, transformed to the rest-frame of \grb.  In both cases, the spectrum exhibits an overall flat shape from $\sim1-1.5$ $\mu$m, with a smooth, gradual upturn at redder wavelengths extending to the edge of our coverage at $\sim2.7$ $\mu$m and a sharp upturn at bluer wavelengths due in part to apparent broad emission features.  The use of different extinction laws and parameters, within the range of uncertainties from our fitting, does not change these fundamental characteristics.

The gradual rise in the spectrum at wavelengths $\lambda \gtrsim1.5$ $\mu$m strongly resembles a power-law shape and therefore this region likely contains a significant contribution from the afterglow of \grb.  In addition, our photometric observation in the F444W filter (which lies redward of our spectroscopic coverage) indicates that the flux continues to rise at longer wavelengths ($\gtrsim3.8$ $\mu$m, rest frame).  The fluxes measured in the F200W, F277W, and F444W filters are consistent with a single power law with an index of $\beta = -0.64\pm0.10$.  The shape of the NIRSpec spectrum at wavelengths $\lambda \gtrsim1.5$ $\mu$m is slightly steeper than this slope, with a power-law index of $\beta = -0.76\pm0.07$, though consistent within the uncertainties. 

At $\lambda \lesssim1.5$ $\mu$m, the spectrum clearly deviates from an extrapolation of the power law at $\lambda \gtrsim1.5$ $\mu$m, exhibiting an overall flat shape and several broad SN-like emission features.  Indeed, we identify two broad emission features located at wavelengths of $\approx0.86$ and $\approx0.92$ $\mu$m, consistent with the Ca\,II NIR triplet and O\,I, respectively. These are typical nebular phase emission lines observed in core-collapse SNe (e.g., \cite{Shahbandeh2022}). We show a zoomed-in comparison of these features with those seen in SN\,1998bw, SN\,2013ge, and SN\,2014ad in Extended Data Figure \ref{fig:CaII}. In addition to the flat spectral shape at $\sim1-1.5$ $\mu$m, these emission features strongly support the identification of SN emission in our spectrum of \grb.  While disentangling the SN and afterglow components is not straightforward, the relative featureless nature of the red end of the spectrum indicates the afterglow component is sufficiently bright to not only affect the overall shape but also dilute SN features with respect to the continuum in that region (see Methods and Extended Data Figure \ref{fig:speccomp_almaxrt} for comparisons with previous SNe). Our observation, therefore, represents the latest NIR spectrum of a SN associated with a GRB to date.

\subsection{Isolating the SN Signal} 
\label{sec:comps}

To separate the afterglow and SN components, we consider several afterglow models.  First we use ALMA and \textit{Swift}/XRT observations obtained at roughly the same phase as our NIRSpec spectrum and model the afterglow at NIR wavelengths as a power law connecting the radio and X-ray data.  We find $F_{\nu} \propto^{-0.63 \pm 0.03}$ (see Methods for details).  We show this power law, normalized to the measured radio/X-ray flux, compared to our spectrum in Figure \ref{fig:specphot}.  While the ALMA-XRT power-law slope is similar to the shape of our spectrum at $\lambda\gtrsim1.5$ $\mu$m, our data is systematically offset to higher flux, indicating the ALMA-XRT power law does not fully capture the afterglow contribution at NIR wavelengths.  Moreover, the implied SN component deviates from the expected spectral shape of a SN (see Methods and Extended Data Figure \ref{fig:speccomp_almaxrt}).

Next we model the afterglow from our spectrum itself, namely as a power law with a slope determined from fitting our spectrum at wavelengths $\lambda\gtrsim1.5$ $\mu$m where the afterglow is likely dominating.  We find a best-fit power law of $F_{\nu}\propto\nu^{-0.76 \pm 0.07}$.  This is steeper than the ALMA-XRT power law, further confirming that interpolating the millimeter and X-ray bands likely does not provide the best representation of the afterglow at these wavelengths.  We then perform a joint fit of a SN template and the fitted power law, with the power-law slope fixed, to determine the best-fit combination of SN and afterglow.  For the SN template we use the $+51$ day spectrum of SN\,1998bw as this is the latest available NIR spectrum of another GRB-SN \cite{Patat2001}, allowing the overall flux normalization to vary.

In Figure \ref{fig:speccomp2} we present the best-fit SN\,1998bw+afterglow spectrum and our spectrum of \grb\ after subtracting the best-fit afterglow component.  We compare our afterglow-subtracted spectrum with the SN\,1998bw spectrum scaled to the distance of \grb\ and the brightness of SN\,1998bw at the phase of our JWST spectrum using the light curve of SN\,1998bw from \cite{Clocchiatti2011}.  The best-fit SN component is $\sim30$\% fainter than the expected brightness SN\,1998bw would have at this distance and phase.  We also compare with late-time spectra of the SN Ic SN\,2013ge \cite{Drout2016} and the SN Ic-BL SN\,2014ad \cite{Shahbandeh2022}.  To directly compare the shapes and features we scale SN\,2013ge and SN\,2014ad to best match the spectral shape and features at the blue end of the afterglow-subtracted spectrum where the SN component dominates.    

These events provide an excellent visual match to the afterglow-subtracted spectrum, confirming that our estimate of the afterglow contribution is reasonable.  In addition, the inferred ratio of Ca\,II/O\,I is a much better match to the ratios seen in the three comparison objects compared to the case of no afterglow subtraction (see Extended Data Figure \ref{fig:CaII}).  While the width of the Ca\,II emission complex exhibits a better match with SN\,2013ge, the afterglow-subtracted spectrum does not show the same strong absorption seen at $\sim1.1$ $\mu$m in SN\,2013ge, possibly due to the SN associated with \grb\ having a higher ejecta velocity.  SN Ic-BL-like velocities are further supported by the better overall match to SN\,2014ad and SN\,1998bw.  The narrower width of Ca\,II compared to SN\,2014ad and SN\,1998bw may be an artifact of the instrumental response impacting the shape at the blue end of the line.  We also identify evidence for a broad emission feature near $\lambda\approx1.5$ $\mu$m, consistent with the location of the 1.503 $\mu$m line of Mg\,I seen in the comparison objects and large samples of other SNe Ic/Ic-BL \cite{Shahbandeh2022}.    
  
In summary, our spectrum is well fit by a SN and power-law model; we do not require another component to explain the spectrum, although we explore the possibility that the afterglow contribution is lower and whether some of the resulting red excess in such a model (see Methods Section \ref{AGmethods}) could be explained by $r$-process emission in Section \ref{sec:rprocess} and Methods Section \ref{rpmethods}.  Importantly, our afterglow-subtracted spectrum is similar to, though slightly fainter than, the expected flux of SN\,1998bw at the distance of \grb\ and the phase of our observations, suggesting the SN associated with \grb\ produced a similar quantity of $^{56}$Ni.  

\subsection{A Modest Nickel Mass Indicates a Typical GRB-SN}

Estimates for the mass of $^{56}$Ni produced in SN\,1998bw range from $\approx0.3-0.7$ M$_{\odot}$ depending on the models and assumptions used to fit the light curve.  \cite{Maeda2003} considered a two-zone model where $\approx0.44$ M$_{\odot}$ of $^{56}$Ni is contained in an outer high-velocity component that rapidly expands and becomes optically thin, explaining the bright peak luminosity.  An additional $\approx0.12$ M$_{\odot}$ exists in an inner dense low-velocity component that explains the linear nature of the light curve at intermediate phases of $\sim100-200$ days.  

We directly estimate the mass of $^{56}$Ni produced by the SN associated with \grb\ by integrating the afterglow-subtracted spectrum.  We estimate the unobserved flux using SN\,2007gr as a spectral template due to its NIR coverage out to the same phase of our observations.  We find that the wavelength coverage of our NIRSpec spectrum accounts for about 50\% of the total emitted flux.  At the phase of our observations, the luminosity of a nickel-powered SN is dominated by the decay of $^{56}$Co, the daughter isotope of $^{56}$Ni.  Assuming a single component of the ejecta and full gamma-ray trapping, we find $M_{\rm Ni}\approx 0.03$ M$_{\odot}$.  Under a more realistic assumption of gamma-ray leakage, with a timescale of $\approx100$ days for the ejecta to become optically thin to gamma-rays (as inferred for SN\,1998bw), we find $M_{\rm Ni}\approx 0.09$ M$_{\odot}$.

The $^{56}$Ni mass we infer assuming gamma-ray leakage is therefore slightly lower than the mass inferred by \cite{Maeda2003} for the inner dense component of SN\,1998bw, consistent with our inference that the SN associated with \grb\ is slightly fainter than SN\, 1998bw at late time.  Of course, assuming a different afterglow contribution in our spectrum will affect the estimated mass.  Our inferred mass is consistent with the results of \cite{Srinivasaragavan2023}, who found best-fit values from modeling the early light curve of \grb\ in the range $M_{\rm Ni} = 0.05-0.25$ M$_{\odot}$, depending on assumptions about the host extinction, with a 99\% upper limit of $M_{\rm Ni}<0.36$ M$_{\odot}$.  These values are lower than most early light curve estimates for SN\,1998bw.  This may indicate a lower ratio of the outer-to-inner ejecta components compared to SN\,1998bw, or that a two-component model is not needed to explain the SN associated with \grb.  Our results, combined with the early light curve estimates, conclusively rule out the possibility that the SN was unusually bright compared to previous GRB-SNe.  Crucially, our spectroscopic detection of the SN confirms that the marginal deviation from a typical afterglow in the early light curve was indeed due to the SN.

\subsection{No Signs of $r$-Process Enrichment}
\label{sec:rprocess}

The identification of the SN associated with \grb\ allows us to constrain the presence of $r$-process material.  One possibility is that the red excess in our spectrum consists of a combination of afterglow and emission from $r$-process elements.  \cite{siegel2019} outline how a collapsar with a massive transient disk can lead to $r$-process production. However, the observational impact of $r$-process material, if it is produced, is highly dependent on the degree of outward mixing. In particular, \cite{siegel2019} present two toy scenarios: one (the ``MHD" case) in which 0.025 M$_{\odot}$ of $r$-process elements are mixed uniformly throughout the SN ejecta with $v<0.15c$, and one (the ``collapsar" case) in which 0.25 M$_{\odot}$ of $r$-process elements are confined to $v<0.015c$. In both cases, $0.25$ M$_\odot$ of $^{56}$Ni is mixed in the ejecta. In the MHD case, the $r$-process material tracks $^{56}$Ni, while in the collapsar case, the $r$-process elements are embedded behind the $^{56}$Ni. In truth, the degree of mixing in the collapsar wind scenario is unknown, likely variable with progenitor properties, and may be sufficient to mix $r$-process elements with the outer layers.

While the MHD scenario has largely been ruled out by early time observations of previous events, few constraints exist on the collapsar wind scenario due to the lack of late-time NIR spectra of GRB-SNe.  Prior to our NIRSpec spectrum of \grb, the latest NIR spectrum of a GRB-SN was that of SN\,1998bw taken at $+51$ days, which we have shown is an excellent match to our spectrum (Figure \ref{fig:speccomp2}) after subtracting our best-fit afterglow power law.  Here we consider the possibility that our best-fit power law overestimates the afterglow contribution and that our much later spectrum of the SN associated with \grb\ differs from the $+51$ day NIR spectrum of SN\,1998bw due to the presence of $r$-process signatures. 

In Figure \ref{fig:rpspeccomp} we compare our NIRSpec spectrum, with various assumptions about the afterglow contribution, to $r$-process enriched SN models from \cite{siegel2019} (with $r$-process masses up to $0.25$ M$_\odot$).  We compare to models corresponding to a phase of 95 days after explosion, the latest phase available, and shift them to the distance of \grb.  At this phase the MHD SN differs considerably from a SN without $r$-process enrichment, producing strong emission at $\approx 1.8-2.4$ $\mu$m that is clearly not present in our spectrum no matter the assumption on afterglow contribution.  The collapsar wind model, on the other hand, largely shows SN features from non-$r$-process elements, though with enhanced flux near $\approx 2$ $\mu$m compared to what is seen in normal SNe.  

Due to the noise in our spectrum, we are unable to identify individual lines in this region of the spectrum, beyond the likely Mg\,I at $\lambda\approx1.5$ $\mu$m.  However, we compare the overall flux level and find that assuming no afterglow contribution (i.e.~the original unsubtracted spectrum) leads to much higher continuum flux than the collapsar wind model for $\lambda>1$ $\mu$m.  Furthermore, the expected strong nebular SN lines are diluted with respect to the continuum (see Methods), indicating an extra continuum source is present (the afterglow of the GRB).  Assuming the afterglow shape and normalization given by interpolating the contemporaneous ALMA and XRT observations, we find that overall the spectrum is inconsistent with the collapsar wind model.  Given the strong resemblance to previous SNe across the full wavelength range when assuming our best-fit afterglow power-law shape and contribution (Figure \ref{fig:speccomp2}), it is unlikely that flux from $r$-process elements are contributing significantly to our spectrum. Our observation highlights the need for a systematic survey of nebular phase LGRB spectra across a broad range of GRB properties, especially in light of the recent theoretical work which correlates these properties to the degree of $r$-process production \cite{siegel2022super, barnes2023hydrodynamic}. We additionally compare our observations to the broadband color evolution models due to $r$-process enrichment from \cite{BarnesMetzger2022} which further highlights the need for spectroscopy (see Methods Section~\ref{rpmethods}).

\section{Host Galaxy Properties}

\subsection{A very low-metallicity, star-forming galaxy}

The host galaxy of \grb\ is readily apparent in our JWST/NIRCam imaging shown in Figure \ref{fig:images}.  Consistent with analysis of the optical HST images \cite{levanboat}, we find that \grb\ is located $0.24\pm0.01$ arcsec ($0.66\pm0.02$ kpc) from the center of its host galaxy, which appears to be a near edge-on system.  From our GALFIT modeling (see Methods) we find that this galaxy is well-described by a single S\'ersic component with index $n=1.2\pm0.1$ and effective radius $r_{e}=2.15\pm0.07$ kpc.  These values represent the mean and standard deviation across the four filters.  The AB magnitudes in each filter corresponding to the best-fit GALFIT models are $m_{\rm F115W} = 21.58\pm0.20$ mag, $m_{\rm F200W} = 20.62\pm0.10$ mag, $m_{\rm F277W} = 20.88\pm0.10$ mag, and $m_{\rm F444W} = 21.38\pm0.05$ mag (not corrected for Galactic extinction).     

In Figure \ref{fig:hostspec} we show the global host spectrum (i.e.~including flux from the entire resolved spectral trace) and the spectrum at the position of \grb\ (see Methods for details of the spectral extractions).  We also show the ``host-only" spectrum, which represents an estimate of the host galaxy spectrum excluding the region of the GRB.  Comparing the spectrum at the position of the GRB with the host-only spectrum clearly shows that certain lines, mostly from molecular H$_2$, are much stronger at the position of \grb.

\begin{figure}
\centering
\includegraphics[width=\textwidth]{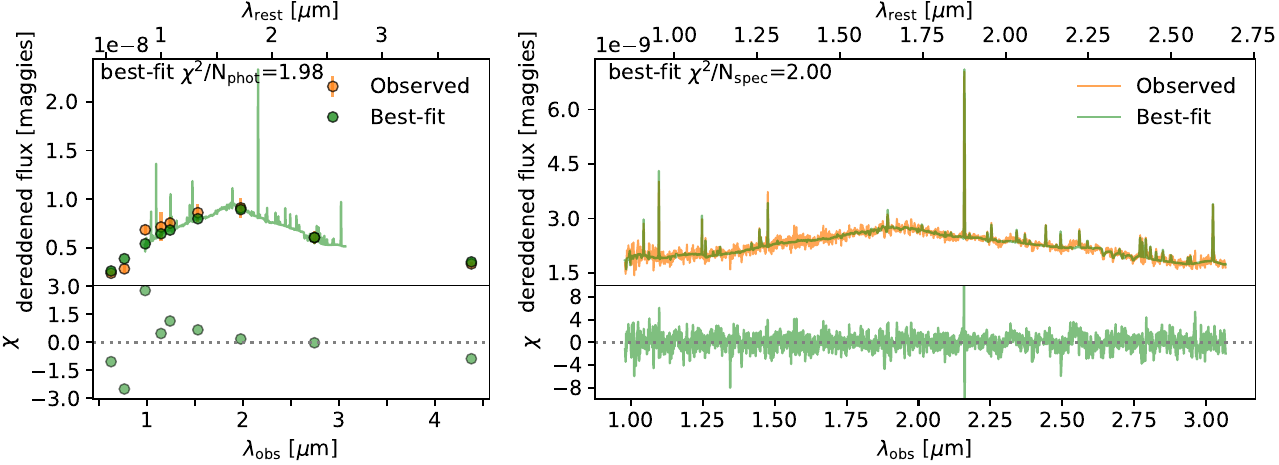}
\caption{Left: Best-fit prospector model photometry (green circles) and spectrum (green) compared to the observed photometry (orange circles; our NIRCam photometry and HST photometry measured by \cite{levanboat}).  Right: Best-fit prospector spectrum compared to our NIRSpec spectrum (orange).  The bottom panels present the residuals.}
\label{fig:prospectorfit}
\end{figure}

We measure the global host properties by fitting the global host spectrum, as well as our NIRCam photometry and HST photometry from \cite{levanboat}, using the stellar population modelling code \texttt{Prospector} \citep{prospector} (see Methods for details of the modeling procedure).  The best-fit model spectrum and photometry compared to the observed data are shown in Figure \ref{fig:prospectorfit}.  We find that the host has a stellar mass of  $\log(M/$M$_{\odot})=9.57^{+0.04}_{-0.06}$ and low stellar and gas-phase metallicities of  $\log(Z_*/Z_\odot)=-0.84\pm0.04$ and $\log(Z_{\rm gas}/Z_\odot)=-0.92^{+0.05}_{-0.04}$, respectively. This is one of the lowest metallicity environments of any LGRB, a class of objects which prefer low-metallicity galaxies \citep{levesque2010host,GrahamFruchter2013,Perley2016review, Graham2019}, and it is, to our knowledge, the lowest metallicity environment of a GRB-SN to date. This may suggest very low metallicity is required to produce a very energetic GRB.  In addition, the galaxy exhibits a recent star formation rate of SFR$_{100\rm Myr} = 0.17 $ M$_{\odot}$/yr.  We also find that the galaxy exhibits a visual extinction of $A_{V} = 0.67^{+0.11}_{-0.07}$ mag.  This is consistent with our extinction constraints from the early-phase JWST data (see Methods) where we found a best-fit total extinction of $A_{V} = 4.63^{+0.13}_{-0.64}$ mag, which is in good agreement, within uncertainties, with the nominal Milky Way value plus the host galaxy extinction found here.  Our SFR and host extinction values are consistent with those measured from H$\alpha$ and Pa$\alpha$ detected in an early-phase X-shooter spectrum of \grb\ \cite{Malesani2023}.  

We additionally model the spectrum \textit{at the site of the GRB}, finding a similar gas-phase metallicity of $\log(Z_{\rm gas}/Z_\odot)=-0.89^{+0.95}_{-0.06}$ and a lower stellar metallicity of $\log(Z_*/Z_\odot)=-1.51^{+0.12}_{-0.32}$ compared to the global host galaxy.  While the upper 1$\sigma$ bound on $Z_{\rm gas}$ extends to solar metallicity due to an isolated region of the posterior, most of the posterior probability is located at low metallicity (see Extended Data Figure \ref{fig:cornerplot}), making it likely that the progenitor of \grb\ originated from a low metallicity environment.

\subsection{Strong H$_2$ Emission at the Explosion Site}

\begin{figure}
\centering
\includegraphics[width=\textwidth]{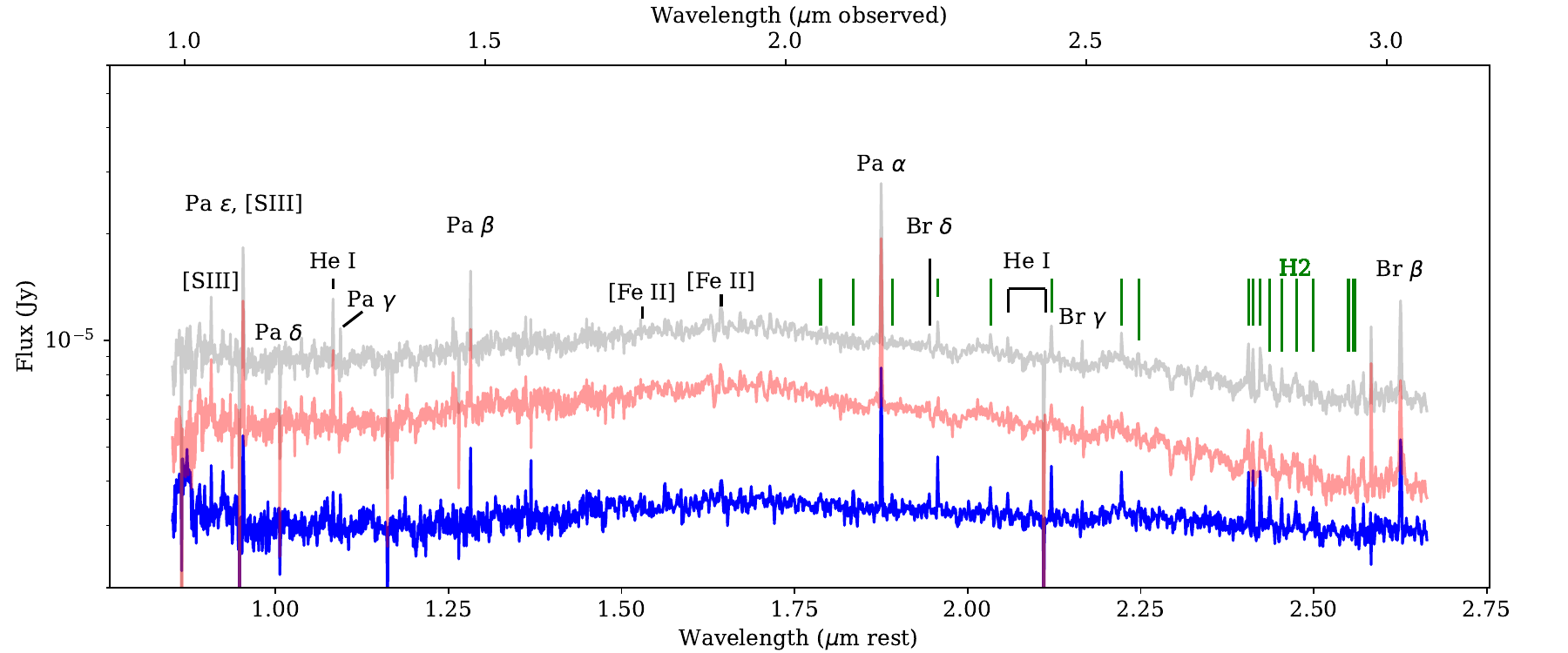}
\caption{Spectrum of the total host galaxy including the site of the GRB (grey), the ``host-only" spectrum excluding the GRB site (red; see Methods Section \ref{sec:obs}), and a narrow aperture centered on the location of \grb\ (blue).  We detect narrow H, H$_2$, He, Fe, and S emission lines from the galaxy.  Importantly, we see that some narrow emisison lines change in strength over this galaxy, having notably strong molecular H$_2$ emission in the region of \grb.  The continuum of the afterglow and SN component can clearly be seen in the blue spectrum as a deviation from the host-only spectrum.}
\label{fig:hostspec}
\end{figure}

We observe many narrow H$_2$ vibrational and rotational emission lines that appear strongest at the site of \grb, as highlighted in Figure \ref{fig:hostspec}. Molecular hydrogen traces dense star-forming regions, consistent with a birth cloud of a massive stellar progenitor of a LGRB. Neglecting the afterglow itself, H$_2$ can be excited by both shocks (driven by e.g., stellar winds or Herbig-Haro objects) or directly by fluorescence \cite{black1987fluorescent, vanzi1997infrared}. Following \cite{izotov2011near}, we compare the ratios of H$_2$ lines from $\simeq1.1-2.1$ $\mu$m to various models of fluorescence vs. collisionally excited emission from \cite{black1987fluorescent} using a simple chi-squared metric (with appropriately propagated uncertainties). Due to the strong detection of many lines (which are predicted to be absent in the case of collisional excitation), we find a better match to fluorescence models, consistent with the dominant excitation method in many low-metallicity, blue compact dwarf galaxies \cite{izotov2011near}.  Our measured line ratios and predicted model ratios are given in Extended Data Table \ref{tab:H2}. 

Only one other LGRB host, that of GRB\,031203 (a relatively faint LGRB), has had a marginal detection of H$_2$ emission \cite{wiersema2018infrared}. Molecular H in \textit{absorption} due to vibrational excitation has also been observed in a small number of events (see, e.g., \cite{prochaska2008first, heintz2019new}). Statistical studies of GRB hosts have found that most lack vibrationally excited H$_2$ (e.g., \cite{tumlinson2007missing}), suggesting that molecular H production is suppressed in LGRB hosts. It has been suggested that this suppression may be in part due to the low metallicities of the hosts \cite{petitjean2006metallicity} or the ongoing star formation, leading to a strong ionizing field \cite{whalen2008molecular}. The low metallicity and modest SFR measured by \texttt{Prospector} suggests that the latter may lead to observable H$_2$ emission in this event. These observations highlight the importance of JWST's unique sensitivity and spatial resolution when analyzing the local environments of LGRB progenitors.

\section{Conclusions}
We present the first detection of a GRB-SN with JWST and the first robust evidence for a SN associated with the highly energetic event \grb. Despite being associated with the brightest GRB ever observed, the SN produced a modest amount ($\approx0.09$ M$_\odot$) of radioactive $^{56}$Ni with no obvious signs of $r$-process nucleosynthesis. The host galaxy suggests a very low metallicity progenitor system -- one of the lowest metallicity environments of all known LGRBs. In addition, the exceptional sensitivity and spatial resolution of JWST allows us to detect, for the first time, a series of multiple molecular H$_2$ emission lines at the position of the GRB -- an observation long anticipated. A secondary site of $r$-process nucleosynthesis remains an open question, which can observationally be uniquely probed by late-time IR spectroscopy. Our findings motivate future JWST campaigns to examine the nebular-phase spectra of SNe associated with LGRBs.

\begin{figure}
\centering
\includegraphics[width=\textwidth]{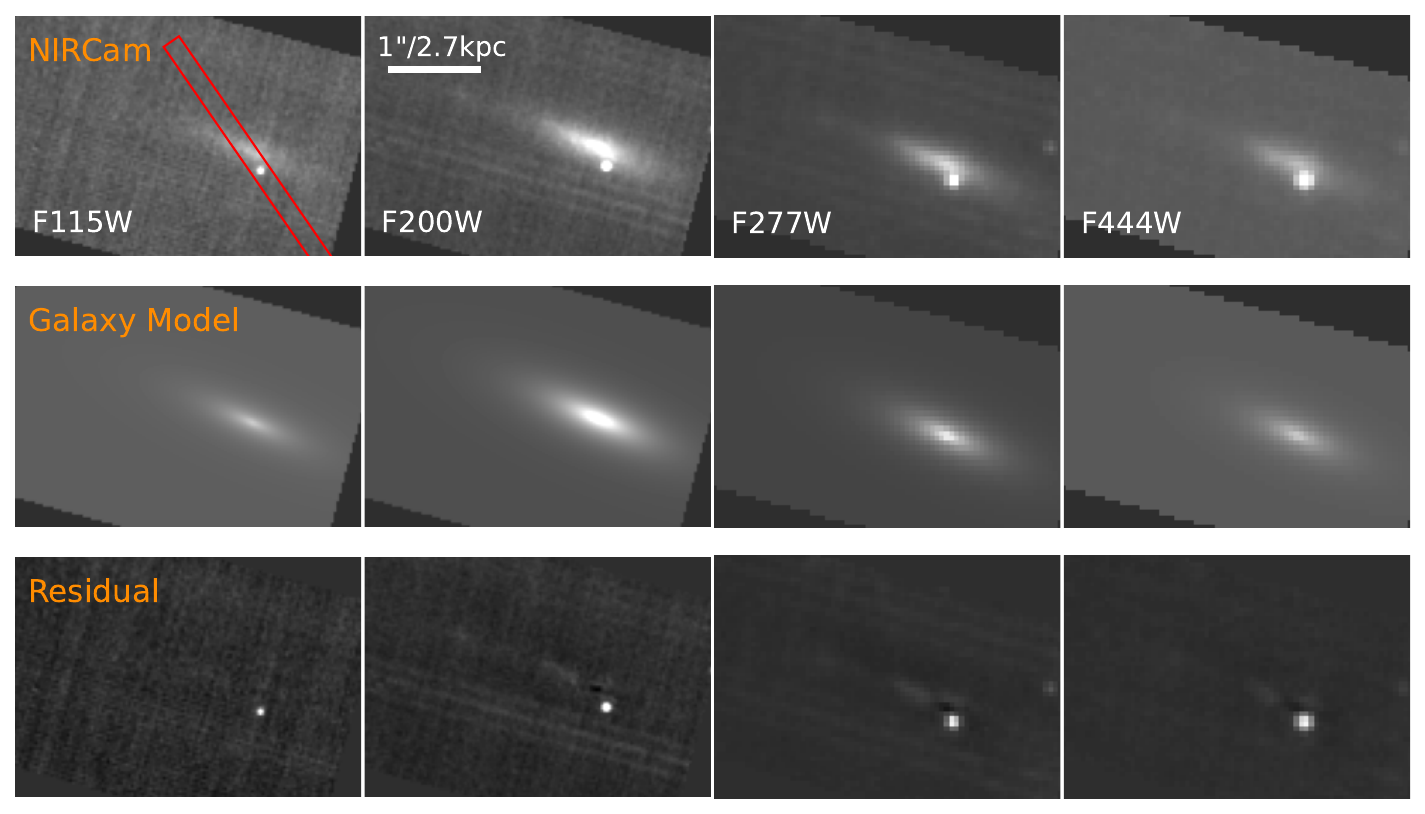}
\caption{Our JWST/NIRCam images of \grb\ (top row), best-fit GALFIT galaxy models (middle row), and GALFIT model subtracted images (bottom row).  Images are shown with North up and East to the left.  A clear point source is detected at the location of \grb.  The red rectangle shows the NIRSpec slit orientation.  PSF photometry of \grb\ was performed on the galaxy-subtracted images.  The host galaxy is well described by a single S\'ersic component, although some residual galaxy structure remains in the F200W, F277W, and F444W filters.}
\label{fig:images}
\end{figure}

\begin{figure}
\centering
\includegraphics[width=\textwidth]{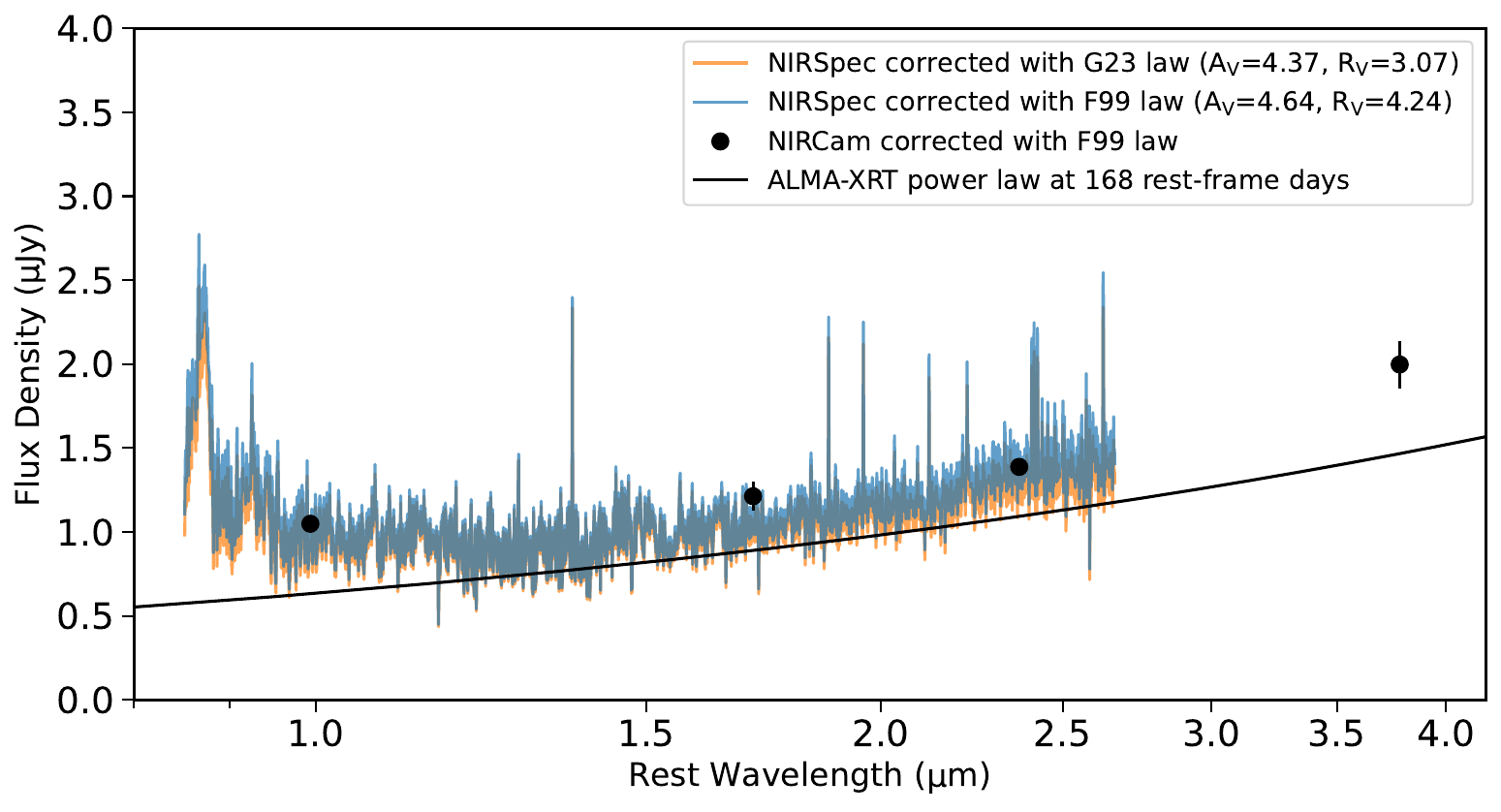}
\caption{Our $+168$ rest-frame phase JWST/NIRSpec G140M+G235M spectrum of \grb\ corrected for extinction (see Methods for spectral extraction details).  We show two versions corrected using the \cite{gordon2023one} (G23; orange) and \cite{Fitzpatrick99} (F99; blue) extinction laws and corresponding best-fit extinction parameters from fitting the early-time NIRSpec/PRISM and MIRI data from \cite{levanboat} as described in Methods Section \ref{sec:dust}.  In both cases, the spectrum appears to exhibit multiple components, with SN-like emission at $\lambda\lesssim1.5$ $\mu$m and rising flux at $\lambda\gtrsim1.5$ $\mu$m likely due to the GRB afterglow power law.  We also show our JWST/NIRCam photometry corrected using a \cite{Fitzpatrick99} extinction law (points), as well as the ALMA-XRT power law (black line).}
\label{fig:specphot}
\end{figure}

\begin{figure}
\centering
\includegraphics[width=\textwidth]{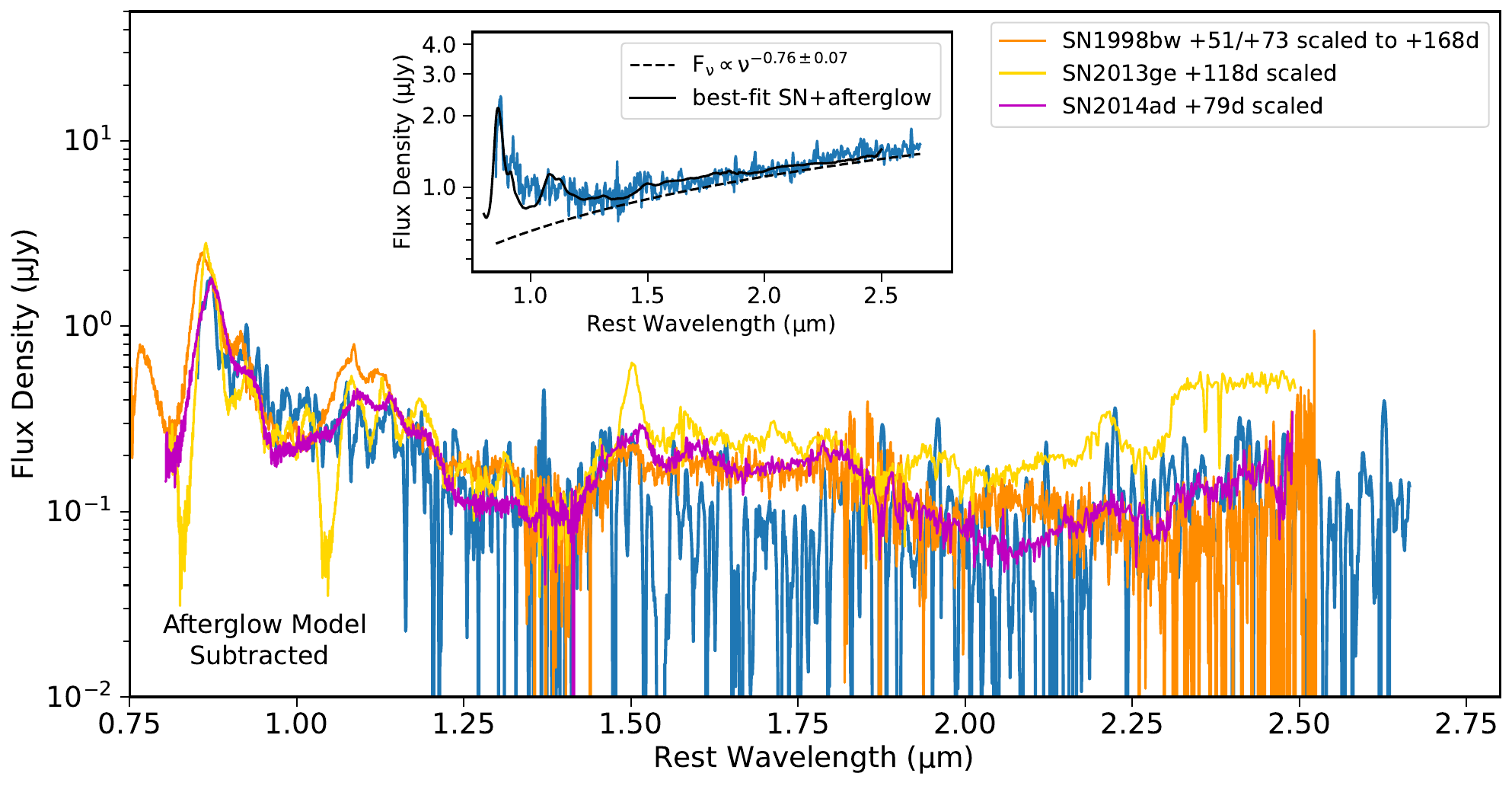}
\caption{Our NIRSpec spectrum of \grb\ (smoothed; blue) after subtracting our best-fit afterglow model.  The unsubtracted spectrum, best-fit afterglow model (dashed black line), and best-fit SN+afterglow (black line) are shown in the inset.  We show late-time spectra of SN\,2013ge (gold; \cite{Drout2016}) and SN\,2014ad (magenta; \cite{Shahbandeh2022}) scaled to match the shape/features of the afterglow-subtracted spectrum at $\lambda\lesssim1.5$ $\mu$m where the SN dominates, demonstrating the overall resemblance with these comparison SNe Ic/Ic-BL.  We also show SN\,1998bw (orange; \cite{Patat2001}) scaled to the distance of \grb\ and the phase of our spectrum, showing that it matches not only the shape but the overall flux level of our spectrum.  The close match with SNe Ic-BL in particular demonstrates the presence of a typical GRB-SN in our spectrum.}
\label{fig:speccomp2}
\end{figure}

\begin{figure}
\centering
\includegraphics[width=\textwidth]{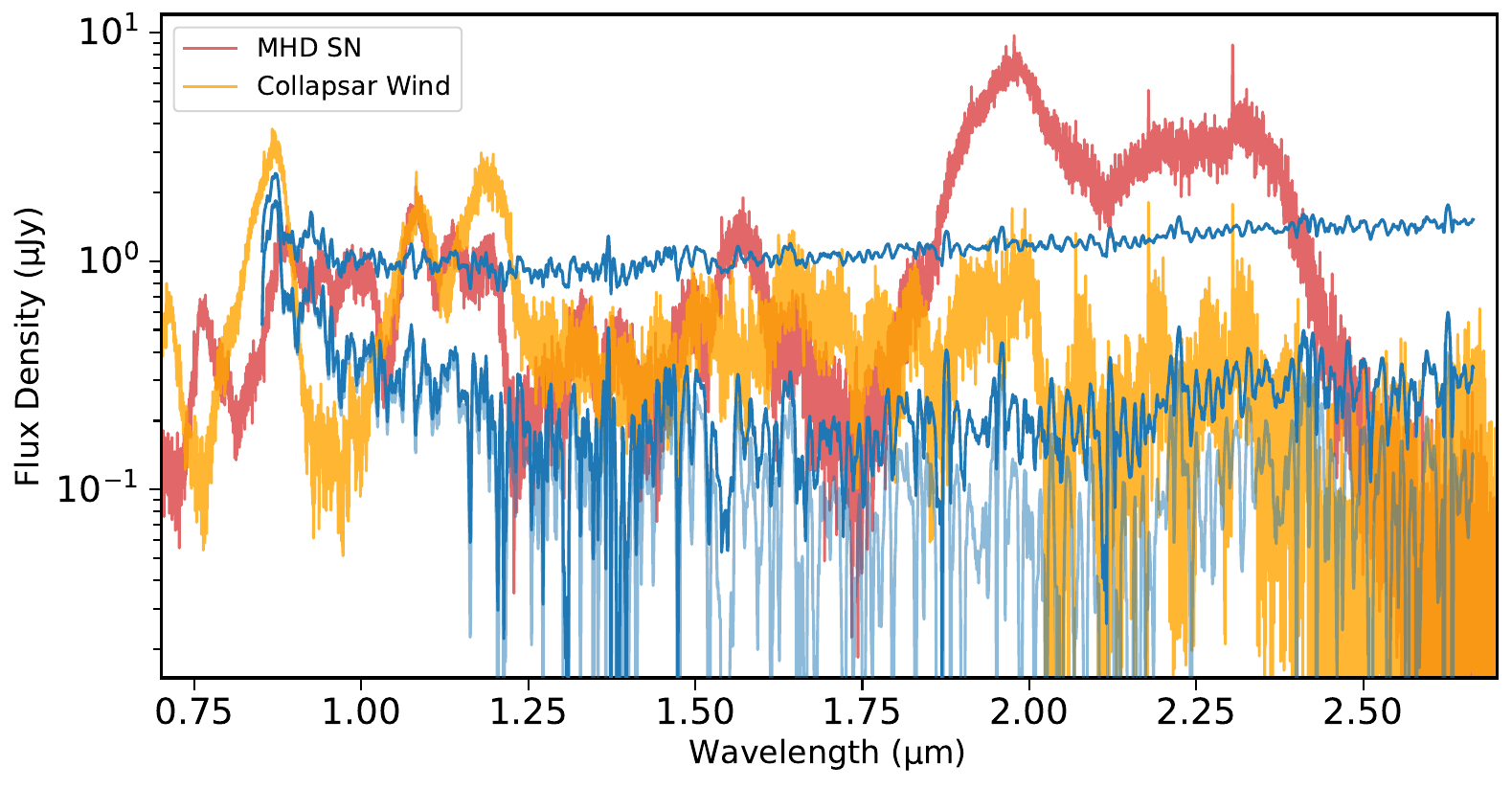}
\caption{Comparison of our NIRSpec spectrum of \grb\ with $r$-process enriched SN models from \cite{siegel2019} corresponding to a phase of 95 days after explosion, the latest phase available.  We show our original spectrum without afterglow subtraction (top blue), as well as the resulting spectra after subtracting the ALMA-XRT power law (middle blue) and our best-fit afterglow model (shown in Figure \ref{fig:speccomp2}; bottom light blue).  Our spectrum, even after accounting for the afterglow, is clearly distinct from the predictions of an MHD SN.  We also do not see evidence for spectral features in our spectrum that can be linked to the collapsar wind model and not attributed to the SN.}
\label{fig:rpspeccomp}
\end{figure}

\newpage

\section{Methods}\label{sec2}

\subsection{Imaging Observations and Photometry}

We obtain imaging of \grb\ with the Near Infrared Camera (NIRCam) using the F115W, F200W, F277W, and F444W filters on 22 April 2023 starting at 07:08 UT.  Each observation consisted of 4 dithered exposures with a total exposure time of 558 seconds.  We downloaded the stage 3 pipeline products from MAST for analysis.  \grb\ is clearly detected, along with its host galaxy.  

To measure the flux from \grb\ in each filter, we first model the host galaxy contribution using the galaxy profile fitting code GALFIT \citep{Peng2010}.  We model the host galaxy as a single S\'ersic component.  During the fit we mask the pixels containing the light from \grb; we fit for transient flux in a later step.  The input, best-fit model, and residual (best-fit model subtracted off) images are shown in Figure \ref{fig:images}.  The residual image for the F115W filter shows no structure indicating the galaxy light is well-described by a single S\'ersic component.  The residual images in the three redder filters, however, exhibit remaining diffuse structure not captured by the model near the center of the galaxy and to the northeast.  While there is no obvious evidence for such diffuse structure emanating from the position of \grb, it is plausible that \grb\ is co-located with a brighter region of its host galaxy that is not captured by our galaxy model.  Such a determination can only be made when the transient fades. 

Next we perform point spread function (PSF) photometry on the residual images at the location of \grb.  Since \texttt{WebbPSF} only generates PSF models for use with stage 2 imaging data, we use the following custom procedure to generate drizzled PSFs for use with stage 3 data.  We generate stage 2 images with model PSFs planted at the location of \grb\ and then run these images through the stage 3 pipeline.  We then use the drizzled PSF models for the PSF fitting of \grb.  We find the following AB magnitudes in each filter: $m_{\rm F115W}=25.10\pm0.05$ mag, $m_{\rm F200W}=24.12\pm0.11$ mag, $m_{\rm F277W}=23.77\pm0.05$ mag, and $m_{\rm F444W}=23.22\pm0.08$ mag (not corrected for extinction which we assess in detail in Section \ref{sec:dust}).  The uncertainties include the systematic uncertainty associated with the GALFIT modeling procedure, which we estimate by comparing PSF photometry with and without galaxy subtraction with GALFIT.

\subsection{Spectroscopic Observations and Data Reduction}

\subsubsection{Late-Time NIRSpec Observations}
\label{sec:obs}
We obtained spectra of \grb\ on 20 April 2023 with the Near Infrared Spectrograph (NIRSpec; \cite{jakobsen2022near}) onboard JWST (program: 2784, PI Blanchard). Our observations began at 14:40 UT, corresponding to a rest-frame phase of 167.7 days since the Fermi GBM trigger.  Spectra were taken with the S200A1 fixed slit and the medium resolution gratings G140M/F100LP and G235M/F170LP, yielding wavelength coverage from $\sim1-3$ $\mu$m.  For each grating/filter setup, we used 5 primary dithers and a total exposure time of 10,942 seconds.  Due to the small offset of \grb\ from its host galaxy \cite{levanboat}, target acquisition was performed using an offset star to ensure proper centering of the source in the slit.

We downloaded and inspected the pipeline products available on the Mikulski Archive for Space Telescopes (MAST).  A resolved trace is clearly present in the individual stage 2 exposures and final combined stage 3 products, indicating a substantial host galaxy contribution.  In addition, a compact trace spanning $\sim$2 pixels is apparent at the red end of the G140M/F100LP spectrum and the G235M/F170LP spectrum at the expected location of \grb\ within the slit.  This trace is also at a consistent offset from the brightest part of the resolved trace representing the center of the host galaxy, confirming that this unresolved trace is the spectrum of \grb.

The pipeline products available on MAST were reduced using nod-subtraction, the default background subtraction method for a point source with multiple dithered exposures.  Due to the resolved nature of the overall trace, we re-reduced the data using the JWST Science Calibration Pipeline with nod-subtraction turned off to reduce the effect of subtracting source flux from itself.  We then extract the 1D spectrum of \grb\ from our re-reduced stage 3 combined and rectified 2D spectra for further analysis.  The final 2D spectra are shown in Extended Data Figures \ref{fig:G140M} and \ref{fig:G235M} (see Figure \ref{fig:images} for the slit orientation).

We use the following extraction procedure to isolate the flux associated with \grb\ from the light of its host galaxy.  We model the spatial profile of the overall trace as a two-component Gaussian with centers fixed at the position of \grb\ and the center of the host galaxy.  We fit the total spatial profile, summed over all wavelengths, to determine the best-fit Gaussian width of each component.  We then fit this model, with widths fixed at these values, to the spatial profile at each wavelength.  We also fit for a linear background component determined from background regions located on both sides of the trace.  The sum of the flux in each fitted Gaussian component thus represent the flux from \grb\ and its host galaxy as a function of wavelength.  The resulting spectra of \grb\ in the G140M and G235M gratings are shown in Extended Data Figures \ref{fig:G140M} and \ref{fig:G235M}, respectively.  The fit to the host galaxy Gaussian component yields a ``host-only" spectrum.   

We note that the background exhibits evidence for PSF artifacts potentially from a nearby bright star (the pseudo-periodic signal at pixel rows $\sim18-24$ in the 2D frames; see Extended Data Figures \ref{fig:G140M} and \ref{fig:G235M}), the most likely explanation given the crowded nature of the field.  Due to the difficulty of accurately modeling this component of the background, our background regions exclude those containing such artifacts.  Our fitted background therefore represents the smooth underlying sky background.  This may mean that the background at the location of \grb\ and its host galaxy is underestimated.  However, we extract regions of the background containing the suspected PSF artifacts from a nearby star and find no evidence that these features are present in our extracted spectrum of \grb.  In addition, the flux from these features decreases towards the spatial location of the GRB spectral trace.

\subsubsection{Combined G140M+G235M Spectrum Compared to Photometry}

Since the photometry was obtained only two days after our NIRSpec spectra, we use the photometry to check the flux calibration of the spectra.  We find that the fluxes in F200W and F277W filters are an excellent match to the flux calibration of the G235M spectrum, whereas the G140M spectrum is offset to higher flux compared to that measured from the F115W filter.  In addition, the G140M is also offset higher than the flux measured in G235M in the wavelength region where the gratings overlap, indicating a systematic offset in the flux calibration between the two gratings.  After correcting the G140M spectrum by a factor of 0.83, the two spectra agree in the overlap region and exhibit an excellent match to the photometry. In Figure \ref{fig:specphot} we show the fluxes in each filter compared with the combined G140M+G235M spectrum after making this correction.  

\subsubsection{Host Galaxy Spectral Extractions}

As seen in Extended Data Figures \ref{fig:G140M} and \ref{fig:G235M}, the host galaxy is resolved in our JWST/NIRSpec observations, extending across $\simeq 10$ rows in our 2D spectra and with numerous narrow emission lines.  To study the global host properties we extract the entire trace including the position of the GRB.  We note that there is significant variation of the strength of some emission lines across the spatial extent of the galaxy, where several lines are stronger at the position of \grb.  To identify these lines and assess any potential variation in the galaxy properties at the position of the GRB, we extract a narrow aperture centered on the position of \grb.  This differs from the Gaussian decomposition procedure described in Section \ref{sec:obs} used to isolate the GRB spectrum as here we are not modeling and subtracting the underlying host spectrum; the goal here is to measure the host properties at the position of the GRB.

\subsubsection{Archival NIRSpec/MIRI Observations}
We obtain archival spectroscopic observations of \grb\ from JWST, observed with NIRSpec and the Mid-Infrared Instrument (MIRI) on 22 Oct 2022 (program: 2782; PI: Levan and originally presented in \cite{levanboat}). These observations correspond to 13.16 and 13.2 days post-burst, respectively. 

At this epoch, the NIRSpec observations were taken in the low resolution PRISM mode, with spectral coverage from $\simeq0.5-5.5$ $\mu$m. The pipeline products from MAST reveal a clear, high SNR trace in the 2D spectrum. The stage 3, reduced spectrum is consistent with that published in \cite{levanboat}, and we thus use it for analysis in this work without additional reductions. 

The MIRI spectrum was taken in the low-resolution spectroscopy (LRS) mode with the P750L disperser. The automatic reduction of the MIRI spectrum failed, likely due to improper selection of the afterglow trace. We use official MIRI reduction pipeline to manually extract the spectrum from the stage 2 products, carefully selecting the correct trace and appropriate background from the nodded 2D image. At this epoch, the afterglow trace is clearly identified in the 2D spectrum and easily isolated using a simple boxcar extraction. We note that MIRI is uncalibrated below $\lambda\lesssim4.5 \mu$m at the time of analysis, and we therefore remove data below this wavelength of the spectrum from analysis. The MIRI observations are qualitatively consistent with that of \cite{levanboat} and well-matched to their near-simultaneous photometric observation in F560W.

\subsection{ALMA Observations}

Following the seven epochs of Atacama Large Millimeter/Sub-millimeter Array (ALMA) observations of \grb\ through program 2022.1.01433.T (PI: Laskar), we obtained two additional epochs with the same program on 01 March 2023 at a mean time of 15:41 UT and on 11 April 2023 at a mean time of 07:55 UT, corresponding to 143.6514 and 183.7838~days in the observer frame, respectively ($\approx124.80$ and $\approx159.67$~days in the rest frame). Both observations utilized two 4 GHz wide base-bands centered at 91.5 and 103.5 GHz, respectively with J1924-2914 as bandpass calibrator and J1914+1636 as complex gain calibrator. The mm-band afterglow previously reported in \cite{Laskar2023} is clearly detected in the pipeline-processed science-ready data products in the first of the two epochs and more weakly ($\approx4.7\sigma$) detected in the second epoch. We performed photometry using \texttt{imfit} in the Common Astronomy Software Applications (CASA; \cite{CASA2022}) and find a best-fit flux density in the two epochs of $(163\pm22)\,\mu$Jy and $(104\pm23)\,\mu$Jy (including a 5\% systematic flux calibration uncertainty) at a mean frequency of 97.5~GHz, along with a position of R.A.$=$19h13m03.50s and decl.$=$19$^\circ$46$^\prime$24.3$^{\prime\prime}$ with an uncertainty of $0.1^{\prime\prime}$ in each coordinate (consistent across both epochs). Together with the last ALMA 97.5~GHz measurement reported in \citep{Laskar2023}, the temporal decline rate of the mm-band afterglow at $\approx99$--188~days after the burst (observer frame) is $\alpha_{\rm mm}=-1.54\pm0.08$, implying an extrapolated mm-band flux density of $(99\pm23)\,\mu$Jy at the time of the NIRSPEC observations (194 days, observer frame). 

\subsection{\textit{Swift}/XRT Observations}

We downloaded the count-rate light curve of the X-ray afterglow of \grb\ from the \textit{Swift}/XRT website\footnote{http://www.swift.ac.uk/xrt curves/01126853}. Using the spectral parameters presented in \cite{Laskar2023} (MW absorption of $N_{\rm H,MW}=5.36\times10^{21}\,$cm$^{-2}$, intrinsic absorption of $N_{\rm H,int}=1.35\times10^{22}\,$cm$^{-2}$, and photon index of $\Gamma_{\rm X} = 1.8566$), we converted the observed count rate to a flux density ($F_{\rm X}$) at 1\,keV and obtained $F_{\rm X}=(9.3\pm3.5)\times10^{-3}\,\mu$Jy at $196^{+5.5}_{-9.1}$ days (observer frame; corresponding to $170.4^{+4.9}_{-7.9}$~days, rest frame) after the burst. Comparing with our ALMA observations, we find that the spectral index between the ALMA and XRT observations at the time of the NIRSpec observations ($\approx194$~days, observer frame) is $\beta_{\rm ALMA-XRT}=0.63\pm0.03$. We refer to this number elsewhere in the text as the ALMA-XRT power law, anchored to the inferred ALMA flux density of $\approx0.1$~mJy at the time of the NIRSpec observations. 

\subsection{Constraints on Foreground Dust from Early-time Spectroscopy}
\label{sec:dust}
Given the location of \grb\ in the Galactic plane ($b\simeq4^\circ$), we expect significant extinction due to interstellar dust in the Milky Way (MW). \cite{schlafly2011measuring} estimates the MW extinction contribution to be $A_V=4.1034\pm0.0553$, assuming the standard extinction factor $R_V=3.1$. As noted in \cite{kann2023grandma}, these dust maps can be unreliable for low galactic latitudes (see \cite{popowski2003large}). Furthermore, this measurement neglects host contribution; however, given the relatively low redshift, we do not expect to easily distinguish between dust arising from either the MW or host galaxy (see the results of e.g., \cite{Malesani2023}). For simplicity, we neglect redshift dependence of dust.

Given the significant uncertainties expected, we opt to use the first epoch of NIRSpec/MIRI data to determine the appropriate extinction correction. We assume that the spectrum is dominated by some unknown combination of an afterglow (power-law model) and a thermal, SN-like component. Unless $r$-process material is significantly throughout within the ejecta, we do not expect a red thermal component at early times. As such, we assume that the event is dominated by a power-law afterglow at $\lambda_\mathrm{obs}\gtrsim 3\mu$m; below this wavelength, it is reasonable that a SN~1998bw-like event could contribute significant flux. We explore the systematic uncertainties associated with the extinction laws and assumptions on the SN contribution.

Few extinction laws are calibrated across the full wavelength range covered by the NIRSpec/MIRI observations. \cite{gordon2023one} recently presented a new extinction law describing $A(\lambda)/A(V)$ as a function of $R(V)$ from $\simeq0.1-30 \mu$m. We contrast this solution with the commonly used extinction law described in \cite{Fitzpatrick99} to quantify systematic uncertainty from assumptions of dust laws. Given a prescribed dust law, we simultaneously fit the observed day 13.2 (observer frame) spectrum to a power law ($F_\nu\propto\nu^{-\beta}$) and extinction parameters $A_V$ and $R_V$ using a Markov-chain Monte Carlo sampler implemented in \texttt{emcee} \citep{foreman2013emcee}. Our models have four free parameters: the overall power-law normalization (``norm"), the power-law index $\beta$, the dust $A_V$ and $R_V$ values and a white noise scatter term. The scatter quantifies the uncertainty in JWST flux estimates as a fraction of the flux. We assume a wide uniform prior for all parameters except normalization, in which we assume a log-uniform prior. 

We first fit using the dust law presented in \cite{gordon2023one}. Fitting all observed wavelengths $\lambda_\mathrm{obs}<8\mu$m, we find $\beta=0.39\pm0.01$, $A_V=4.37\pm0.05$, and $R_V=3.07^{+0.04}_{-0.05}$. At $\lambda<2\mu$m, we find that the residuals are consistent with 0, suggesting no contribution from an additional thermal component. We next exclude wavelengths $<2\mu$m in the fitting process to test the possibility of contamination from either a SN-like or $r$-process thermal event. We find that when excluding these wavelengths, the afterglow model \textit{over}estimates the blue flux.

Next, we fit using the extinction law described in \cite{Fitzpatrick99} (i.e., following the original analysis of \cite{levanboat}). We again emphasize that this extinction law is not calibrated for IR observations and simply extrapolates at these wavelengths. We again simultaneously fit the observed spectrum ($\lambda_\mathrm{obs}<8\mu$m) to a power law and extinction model. We find $\beta=0.41\pm0.01$, $A_V=4.63^{+0.13}_{-0.64}$, and $R_V=4.24^{+0.74}_{-0.64}$. This is significantly different ($>3\sigma$) from the results presented in \cite{levanboat} when only accounting for statistical uncertainties, which we attribute to a tight prior (vs. our flat prior) set by those authors. 

We report the results of our fits in Extended Data Table \ref{tab:dust} and show these data, models and associated residuals in Extended Data Figure \ref{fig:prism}. The residuals of both dust models show significant structure throughout the spectrum. We specifically compare the residuals to a spectrum of SN1998bw taken 12 days post burst and scaled to the redshift of \grb. We note that the statistical uncertainties and systematic difference between these two dust extinction models mean that we are unable to make a conclusive statement on the SN emission from the early-time JWST spectrum. This is a different conclusion from that of \cite{levanboat}, who, given their small \textit{statistical} uncertainties, rule out SN1998bw-like thermal emission at early times.

\subsection{Constraints on the Afterglow Contribution}
\label{AGmethods}

\subsubsection{Initial Comparisons with Previous SNe}

In Extended Data Figure \ref{fig:speccomp_almaxrt} we show our extinction-corrected spectrum (using a \cite{Fitzpatrick99} law and best-fit parameters listed in Extended Data Table \ref{tab:dust}) compared to spectra of SN\,1998bw \cite{Patat2001}, the canonical SN Ic-BL associated with a GRB, and SN\,2013ge \cite{Drout2016}, one of the few SNe Ic with high S/N late-time NIR spectra, taken at $+51$ and $+118$ days after peak, respectively.  To achieve complete overlap with the blue end of our spectrum, we combine the $+51$ day NIR spectrum of SN\,1998bw with an optical spectrum taken at $+73$ days.  We scale the spectra of SN\,1998bw and SN\,2013ge to the distance of \grb\ and use their light curves \cite{Clocchiatti2011,Drout2016} to normalize to their brightnesses at the phase of our \grb\ spectrum.  Our spectrum of \grb\ is brighter than the comparison SNe would be and relatively featureless with a different overall spectral shape, consistent with significant contamination from the afterglow.  Our spectrum exhibits flux increasing at $\lambda\gtrsim1.5$ $\mu$m, whereas the comparison SNe exhibit declining flux.

The emission features shown in Extended Data Figure \ref{fig:CaII} exhibit similar, though slightly narrower, widths than the corresponding features in SN\,1998bw, SN\,2013ge, and SN\,2014ad.  Due to the lack of a late-time light curve for SN\,2014ad, we scale its spectrum to roughly match SN\,1998bw for comparison purposes.  In addition, the lines in our JWST spectrum are diluted in strength and exhibit a different flux ratio.  This, combined with the rising flux to the red, means there is no simple luminosity scaling that will bring our spectrum of \grb\ into agreement with the comparison spectra.  These observations are consistent with afterglow contamination. Furthermore, the lack of many strong SN features other than the two identified indicates that the SN associated with \grb\ is not substantially brighter than SN\,1998bw and SN\,2013ge.

\subsubsection{Constraints from Contemporaneous ALMA and \textit{Swift}/XRT Observations}

Determining the afterglow contribution is critical to constrain the presence of SN emission and a possible contribution from $r$-process material.  First we consider the power law formed by the ALMA and XRT observations that we obtained around the same phase as our JWST observations.  We analyze the residual spectrum by subtracting off the ALMA-XRT power law from our spectrum of \grb\ which we show, compared to SN\,1998bw and SN\,2013ge, in Extended Data Figure \ref{fig:speccomp_almaxrt}.  While the resulting spectrum more closely matches the shape of the SNe compared to the unsubtracted spectrum, especially at the blue end, the shape at $\lambda\gtrsim1.5$ $\mu$m still exhibits rising flux substantially different from the SNe.  Given the lack of strong emission features in this region, the most likely explanation is that the ALMA-XRT power-law model does not adequately capture the afterglow contribution.  In Sections \ref{sec:rprocess} and Methods Section \ref{rpmethods} we consider whether this red excess could be due to emission from $r$-process material.

\subsubsection{Varying the Afterglow Contribution}

Next we consider the best-fit power law from fitting our spectrum at $\lambda\gtrsim1.5$ $\mu$m (shown in Figure \ref{fig:speccomp2}) and analyze how the implied SN component changes with different afterglow normalizations.  We scale the best-fit power law by factors of 0.3, 0.6, 0.9, and 0.95 to generate four potential afterglow models, subtract them from the spectrum, and compare the resulting residual spectra with SN\,1998bw and SN\,2013ge.  In Extended Data Figure \ref{fig:speccomp1} we show the residual spectra and afterglow models for the four scalings.  When scaled by 0.3 and 0.6, the residual spectra still exhibit flux rising to the red, as in the unsubtracted spectrum, indicating that these models likely do not account for all of the afterglow flux. 

In addition, there is a mismatch between the flux ratios of the expected emission lines.  In other words, the detection of the Ca\,II NIR triplet at the strength we see, would imply the detection of other lines at strengths that are not observed.  Of course, this reasoning relies on the assumption that the SN associated with \grb\ should appear similar to previous SNe Ic/Ic-BL.  Indeed it is possible that this SN may not show the same features as previous events, and potentially an additional component from $r$-process emission which we assess in Sections \ref{sec:rprocess} and Methods Section \ref{rpmethods}.  However, the lack of strong lines in this region indicates that the SN associated with \grb\ is likely fainter than these toy models suggest and the afterglow correspondingly brighter (as found when performing a joint SN+afterglow fit; Section \ref{sec:comps}), such that most emission lines are diluted with respect to the continuum and not detectable.

If instead the best-fit power law is scaled by 0.9, the residual spectrum appears consistent with the comparison spectra, and is a close match to the overall flux level of SN\,1998bw.  Note that this is similar to the best-fit scaling (0.93) when performing the joint SN+afterglow fit as described in Section \ref{sec:comps} and shown in Figure \ref{fig:speccomp2}.  Larger afterglow contributions (e.g., scaling by 0.95) yield an overall steeper slope, inconsistent with the comparison objects.

\subsection{Comparison with $r$-Process Light Curve Models}
\label{rpmethods}

We also consider the $r$-process enriched SN light curve models from \cite{BarnesMetzger2022}.  In Extended Data Figure \ref{fig:rpcolors} we show the $J-H$ and $J-K$ color evolution of these models, for a SN Ic-BL with a typical simulated ejecta mass of 3.96 M$_{\odot}$, a $^{56}$Ni mass of 0.33 M$_{\odot}$, an $r$-process material mass of 0.03 M$_{\odot}$ and various levels of $r$-process mixing from no mixing to nearly fully mixed, compared to the colors of the SN component of \grb\ under different afterglow assumptions.  We calculate $J-H$ and $J-K$ colors by convolving the filter bandpasses with our NIRSpec spectrum \textit{after} subtracting the afterglow models.  We show the resulting colors for the afterglow models considered in Section \ref{AGmethods} (the ALMA-XRT power law and the best-fit power law from fitting the red end of our spectrum with various normalizations; see Extended Data Figures \ref{fig:speccomp_almaxrt} and \ref{fig:speccomp1}).

The $J-K$ colors of the afterglow-subtracted spectra match the $r$-process enriched models when scaling the best-fit power law by $\lesssim$0.9.  Decreasing the afterglow contribution leads to more residual red light, leading to redder colors.  When scaling by $\lesssim$0.6, including the ALMA-XRT model, the $J-K$ colors, if reddened due to $r$-process material, would imply near complete mixing.  In this case, strong broad emission lines from $r$-process elements would be expected, as seen in the MHD model in Figure~\ref{fig:rpspeccomp} but not in our data.  In addition, for a given afterglow contribution, the $J-H$ colors imply a different degree of $r$-process mixing than the $J-K$ colors, suggesting the reddening source is not due to $r$-process emission.  

In Extended Data Figure~\ref{fig:rpcolors} we also show the colors of SN\,1998bw and SN\,2013ge calculated from their late-time NIR spectra.  SN\,1998bw is notably blue -- bluer even than the models without $r$-process -- suggesting these models do not fully capture the range of possible spectral energy distributions of typical GRB-SNe.  SN\,2013ge is notably red, consistent with the $r$-process enriched models for a mixing fraction of $\sim$10\%.  This event, however, exhibits a clear example of CO emission increasing the flux in K-band.  These comparisons highlight that, without spectra, other sources of reddening are difficult to disentangle from that due to $r$-process material.  Similar conclusions have been drawn from studies of large samples of SNe Ic-BL light curves \cite{Anand2023}.  We note that the spectrum of \grb\ after subtracting the best-fit power law scaled by 0.9, which yields a good visual match to SN\,1998bw and SN\,2013ge (Extended Data Figure \ref{fig:speccomp1}), exhibits a $J-K$ color that is $\approx0.2$ mag redder than the no $r$-process model.  However, as can be seen in Extended Data Figure \ref{fig:speccomp1}, there is an upturn in the spectrum in K-band at the expected location of first overtone CO emission, similar to that seen in SN\,2013ge.

\subsection{Host Galaxy Modeling}

We use \texttt{Prospector} \citep{prospector}, a Bayesian galaxy SED-fitting code to simultaneously fit the global host galaxy photometry and spectroscopy. Additionally, we fit the spectrum extracted at the position of the GRB to compare the global host properties and those
at the GRB position. We adopt the MIST isochrones \citep{Choi2016}, and the C3K stellar spectral libraries in the Flexible Stellar Population Synthesis (FSPS; \cite{Conroy2009, Conroy&Gunn2010}) framework. The stellar population is described by redshift, stellar mass, velocity dispersion, stellar metallicity, and a step function nonparametric star formation history (SFH) with 14 time bins \citep{leja19}. The nebular emission is parameterized by gas-phase metallicity and ionization parameter using the \cite{Byler2017} CLOUDY grid. We simultaneously fit simple Gaussians to lines that are not included in our emission line model that assumes all the emission is powered by the stars, namely the He\,I, [Fe\,II], and H$_2$ emission lines, with the same kinematics but free amplitudes as our CLOUDY grid. We assume a flexible two-component dust attenuation model accounting for birth cloud and diffuse dust separately \citep{Kriek&Conroy2013}. Variation in the shape of the attenuation curve is enabled using with a power-law modification to a Calzetti curve \citep{Noll2009}. We also incorporate the contribution of dust emission to the infrared photometry using a three-parameter model \citep{Draine&Li2007}. To fit the spectroscopy and photometry together, we marginalize over the shape of the observed spectrum (thereby avoiding any wavelength-dependent flux calibration issues) with a polynomial; in this manner, the normalization and shape of the SED is entirely determined by the photometry, or not constrained at all for the GRB position where there is no photometry. We also include a jitter parameter that inflates the spectroscopy uncertainties to account for imperfect JWST flux calibration and slit losses, finding typical values of $1.5-2$, consistent with other early JWST spectroscopic analyses. Finally, we use a pixel outlier model to downweight pixels not consistent with our model \citep{hogg10}, which are typically identified at a 1-2\% level. In summary, the SED model for the host galaxy fit has 28 free parameters, and the fit to the spectrum at the GRB position has 24 free parameters.

\backmatter





\bmhead{Acknowledgments}

P.K.B.~acknowledges support from a CIERA Postdoctoral Fellowship. V.A.V.~acknowledges support by the NSF through grant AST-2108676. The authors thank O.~Fox and M.~ Shahbandeh for assistance reducing the MIRI spectrum. This study was enabled in part by a Radboud Excellence Fellowship from Radboud University in Nijmegen, Netherlands. B.D.M.~acknowledges support from the National Science Foundation (grant number AST-2002577).  This work is based on observations made with the NASA/ESA/CSA James Webb Space Telescope. The data were obtained from the Mikulski Archive for Space Telescopes at the Space Telescope Science Institute, which is operated by the Association of Universities for Research in Astronomy, Inc., under NASA contract NAS 5-03127 for JWST. These observations are associated with programs 2784 and 2782. This paper makes use of the following ALMA data: ADS/JAO.ALMA\#2022.1.01433.T. ALMA is a partnership of ESO (representing its member states), NSF (USA) and NINS (Japan), together with NRC (Canada), MOST and ASIAA (Taiwan), and KASI (Republic of Korea), in cooperation with the Republic of Chile. The Joint ALMA Observatory is operated by ESO, AUI/NRAO and NAOJ. This work makes use of data supplied by the UK Swift Science Data Centre at the University of Leicester and of data obtained through the High Energy Astrophysics Science Archive Research Center On-line Service, provided by the NASA/Goddard Space Flight Center.

\bmhead{Author Contributions}

PKB led the overall project from data proposal and acquisition to data analysis and the writing of the manuscript.  PKB is the PI of JWST program 2784, the primary data analyzed in this work.  VAV, as co-PI of JWST program 2784, co-conceived of the original JWST proposal.  VAV contributed significant analysis of both our late-time JWST data and the archival JWST data from program 2782 and a significant contribution to the writing of the manuscript.  RC, co-PI of JWST program 2784, made significant contributions to the original JWST proposal and provided significant comments on the analysis and manuscript.  TL analyzed the ALMA and XRT data, wrote the Methods sections describing that data, and provided comments on the manuscript.  YL and JL carried out the \texttt{Prospector} modeling of the host galaxy spectra and photometry and wrote the Methods section about such modeling.  JP performed the PSF fitting of \grb\ in the NIRCam images.  EB and RM provided substantial comments on the manuscript and the original JWST proposal.  KA, YC, and TE assisted with the acquisition of the ALMA and XRT data and provided comments on the manuscript.  JMP assisted with the reduction of the NIRSpec spectroscopy.  DS provided the $r$-process spectral models for comparison and substantial comments on the manuscript.  BM, JB, DK, HS, NL, and SKY provided comments on the manuscript and the original JWST proposal.  AR assisted with the NIRCam photometry and commented on the manuscript.   

\bmhead{Data Availability}

The JWST data analyzed in this work associated with programs 2784 and 2782 are publicly available on the MAST archive.

\bmhead{Code Availability}

The software tools used in this work (JWST Science Calibration Pipeline, GALFIT, \texttt{WebbPSF}, \texttt{Prospector}, and CASA) are publicly available. 












\begin{appendices}

\section{Extended Data}\label{secA1}


\begin{figure}[h]
\centering
\includegraphics[width=\textwidth]{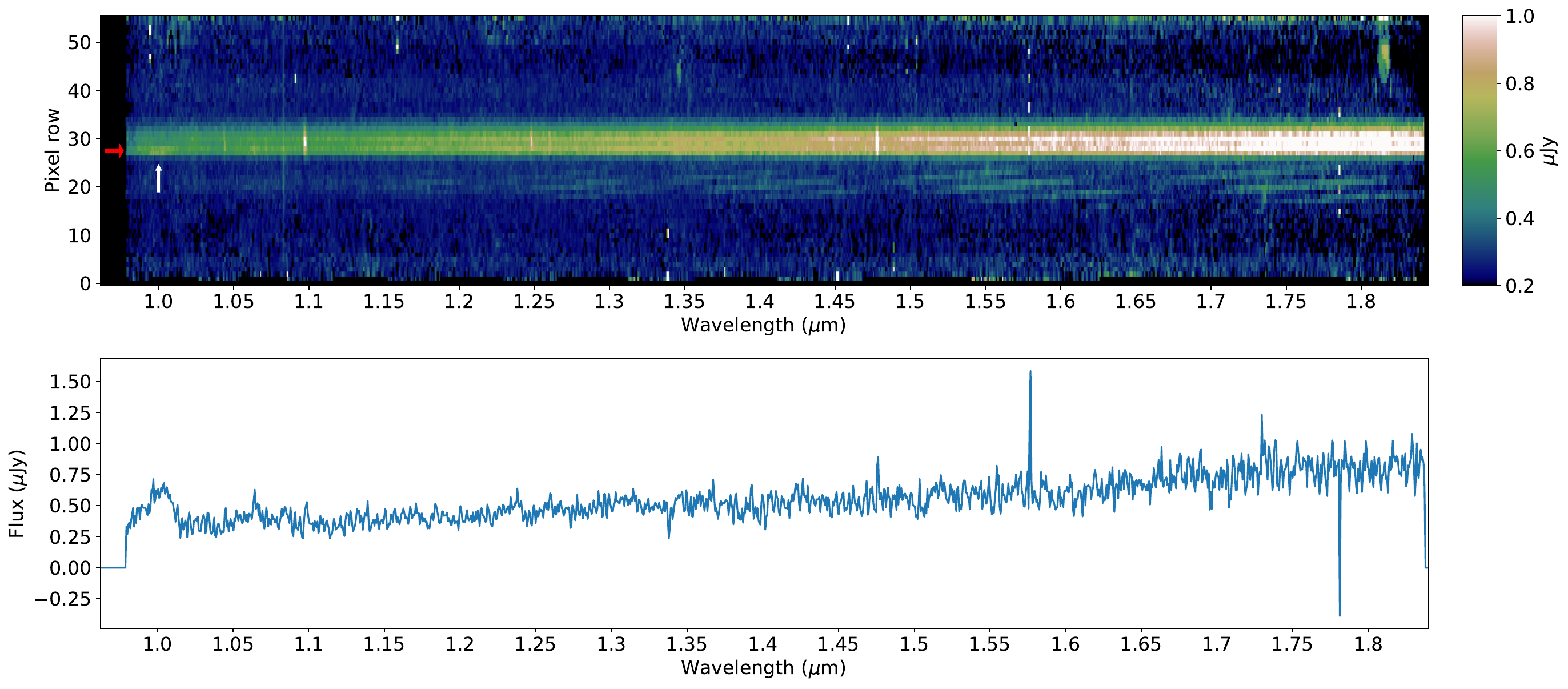}
\caption{Top: Final combined 2D NIRSpec/G140M spectrum of \grb\ resulting from our re-reduction.  The trace is clearly dominated by the spatially resolved host galaxy.  A broad emission feature is visible near $\approx1$ $\mu$m (white arrow) at the expected spatial location of \grb\ (red arrow).  The background below the trace contains structure that is likely due to the diffraction spike of a nearby star.  Spatially resolved emission lines from the host galaxy are also detected.  Bottom:  1D spectrum of \grb\ extracted via a two-component Gaussian fit to the spatially resolved trace to isolate the spectrum of \grb\ from the host galaxy contribution.}
\label{fig:G140M}
\end{figure}

\begin{figure}[h!]
\centering
\includegraphics[width=\textwidth]{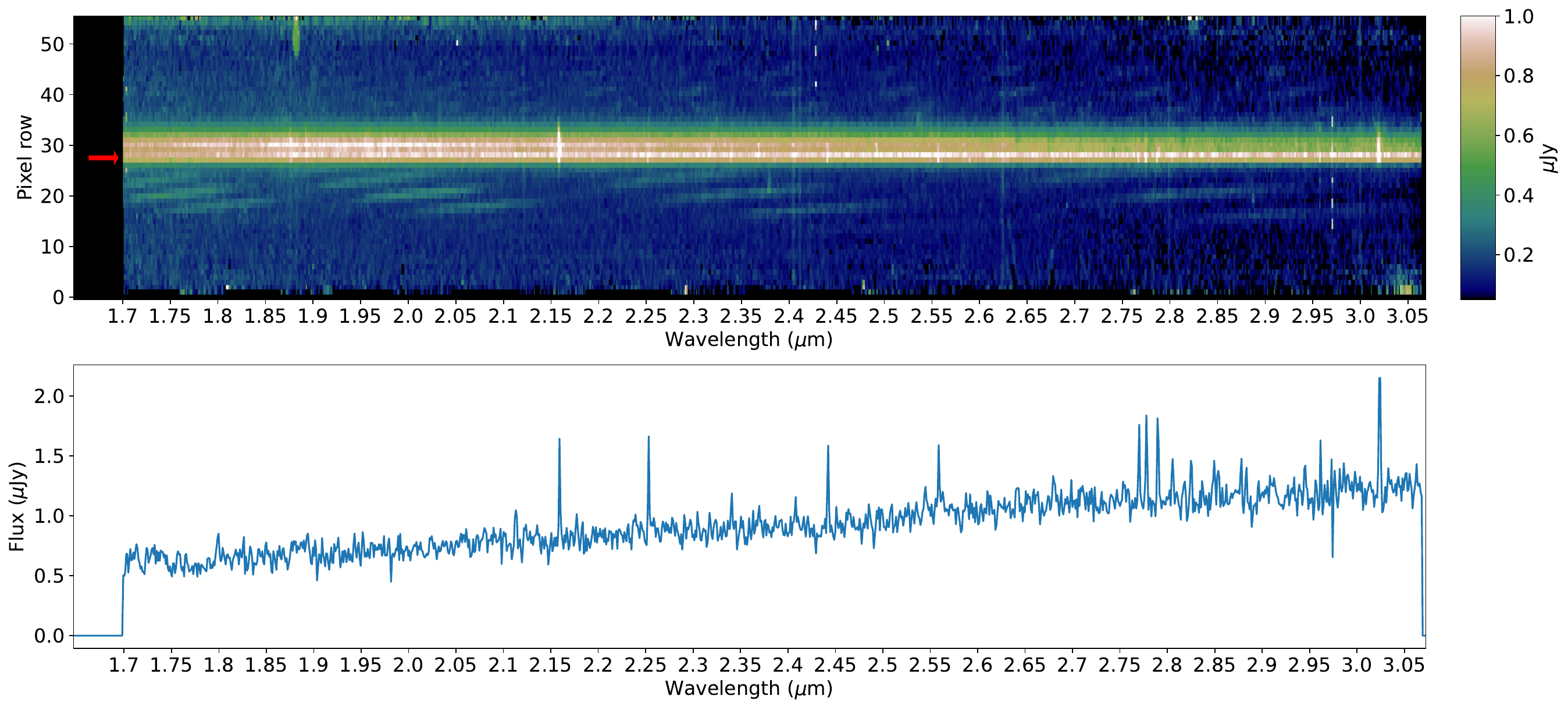}
\caption{Same as Figure \ref{fig:G140M} but for the G235M grating.  The trace from \grb\ (red arrow) is more clearly visible at the red end.  Notably, several host galaxy emission lines appear to be stronger at the position of \grb\ than the rest of the galaxy, resulting in significant excess flux from these lines appearing in the \grb\ spectrum (see Figure \ref{fig:hostspec} for line identifications).}
\label{fig:G235M}
\end{figure}

\begin{figure}[h]
\centering
\includegraphics[width=\textwidth]{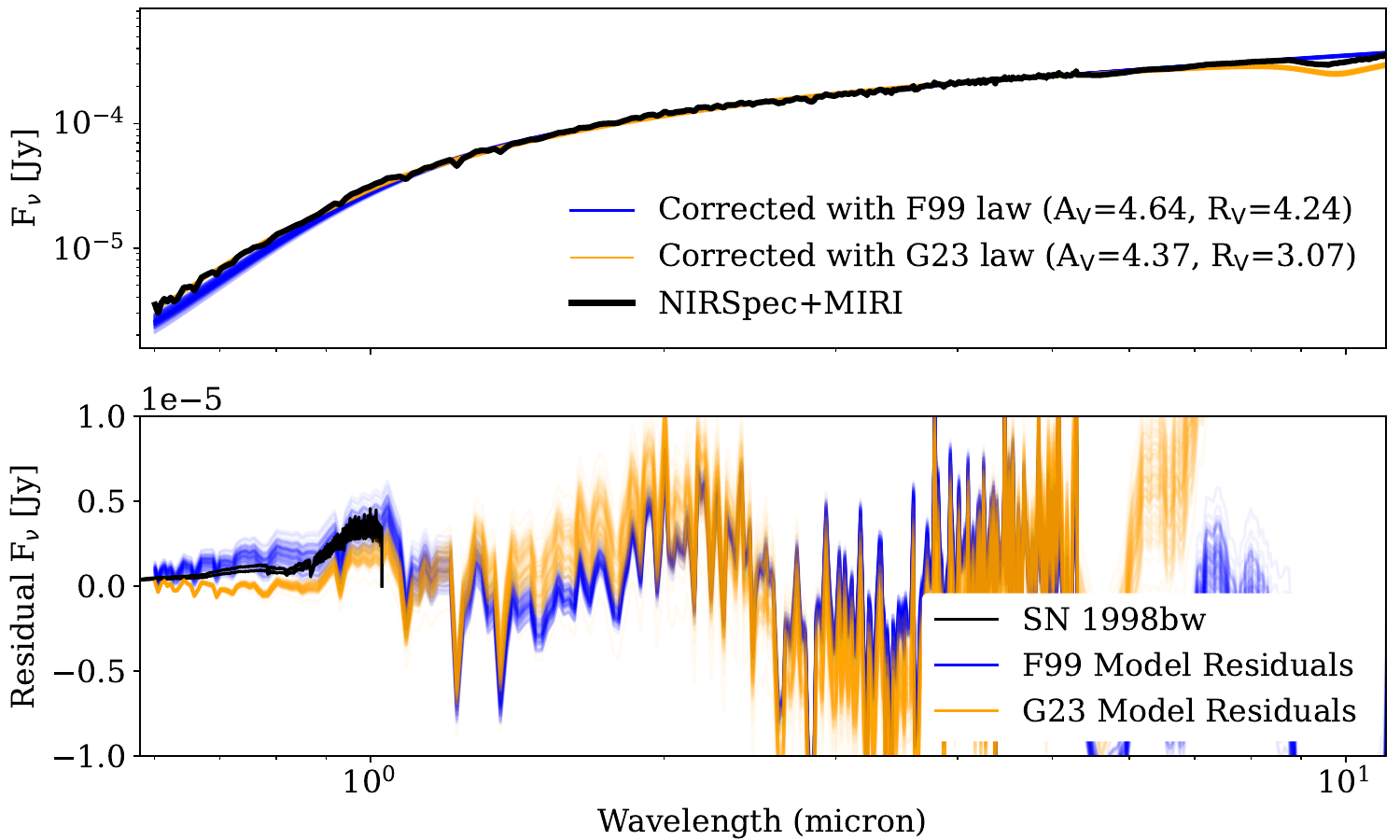}
\caption{Top: Comparison of the early-time NIRSpec/PRISM and MIRI spectra (black) along with two models for the afterglow and dust law (\cite{Fitzpatrick99} in blue, and \cite{gordon2023one} in orange). Each line of the model represents a draw from the posterior. Note that there is a silicate feature at $\simeq10\mu$m which is not properly modeled without a detailed dust composition; here, we fit $\lambda<8\mu$m to avoid this feature. Bottom: Model residuals compared to SN\,1998bw (black; \cite{Patat2001}). Again, individual lines represent independent draws from the model posterior. For both laws, strong systematic residuals are found across the full wavelength range. A SN\,1998bw-like supernova cannot be ruled out.}
\label{fig:prism}
\end{figure}

\begin{figure}[h]
\centering
\includegraphics[width=\textwidth]{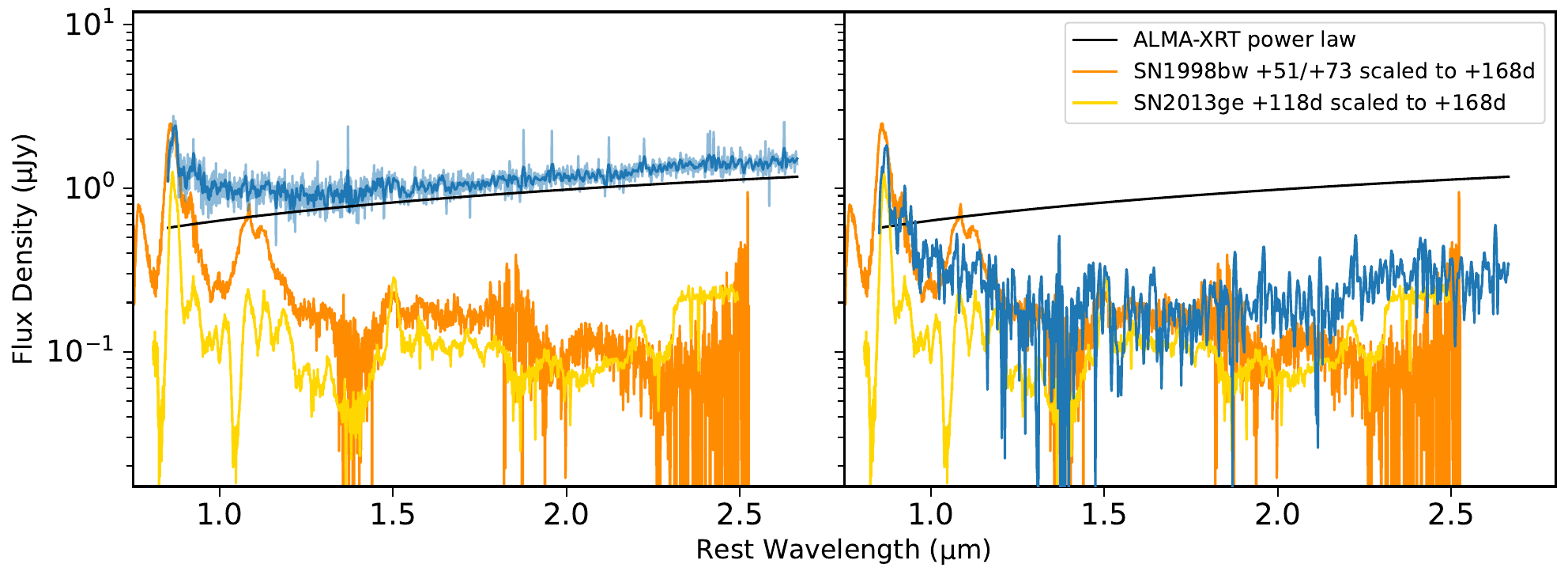}
\caption{Left: Our NIRSpec spectrum of \grb, corrected for extinction using a \cite{Fitzpatrick99} law and best-fit parameters from Section \ref{sec:dust} (blue), compared to ground-based late-time NIR spectra of SN\,1998bw (orange) and SN\,2013ge (gold) scaled to the distance of \grb\ and the same phase of our observations.  Right: Comparison between SN\,1998bw and SN\,2013ge and our spectrum of \grb\ (smoothed) after subtracting an estimate of the afterglow contribution as described by the power law connecting ALMA and \textit{Swift}/XRT observations taken around the same phase as our JWST data ($F_{\nu} \propto \nu^{-0.63}$; black line).  Significant flux rising toward the red remains in the resulting subtracted spectrum, inconsistent with the comparison SNe, indicating this model likely does not adequately describe the afterglow contribution at these wavelengths.}
\label{fig:speccomp_almaxrt}
\end{figure}

\begin{figure}[h]
\centering
\includegraphics[width=\textwidth]{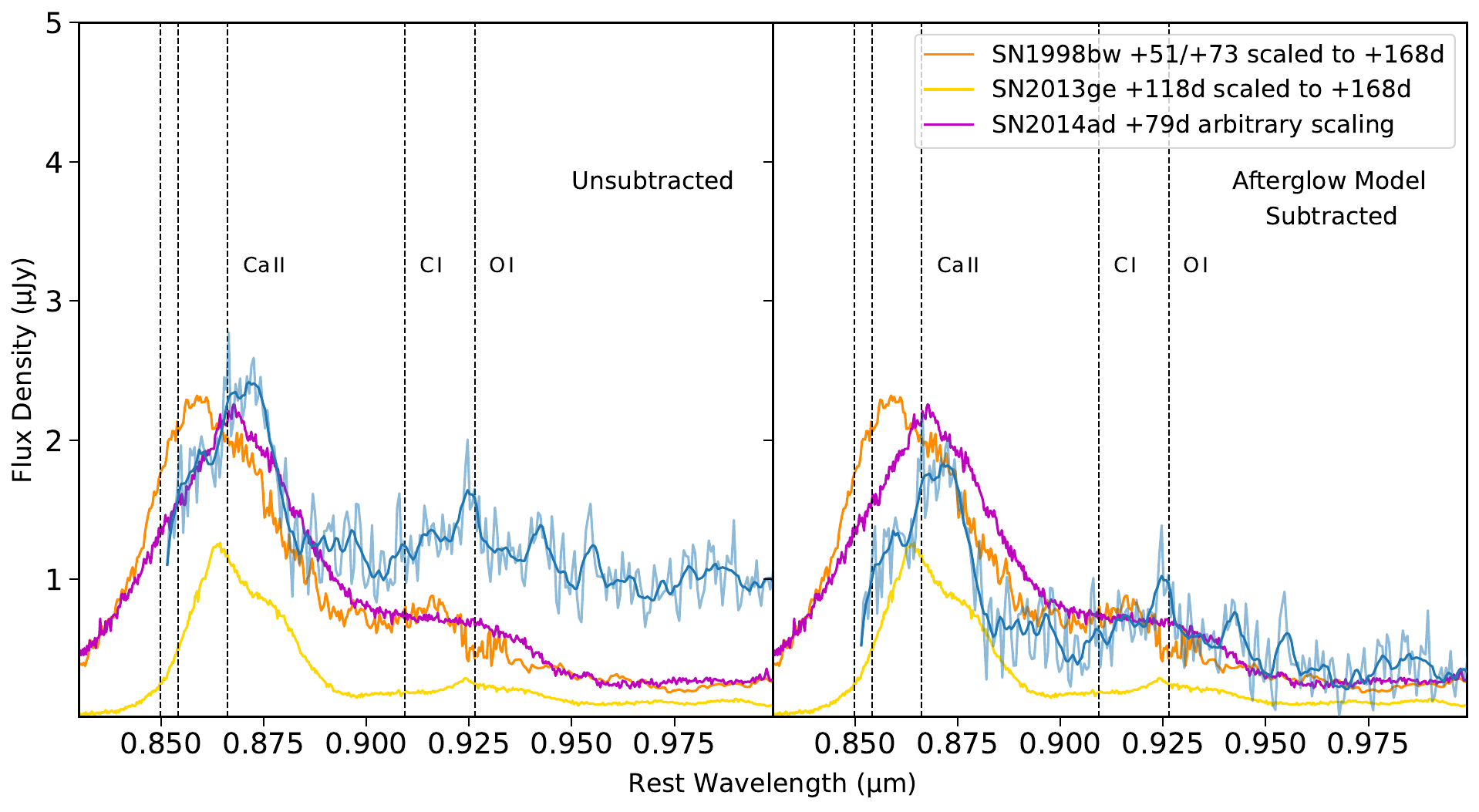}
\caption{Left: Zoom-in on the blue end of our spectrum of \grb\ highlighting the broad emission features we attribute to the Ca\,II NIR triplet and O\,I.  Also shown are comparison spectra of SN\,1998bw (orange) and SN\,2013ge (gold) both scaled to the distance of \grb\ and their brightness at the phase of our observations, as well as SN\,2014ad (magenta) arbitrarily scaled.  Right: Spectrum of \grb\ after subtracting our best-fit afterglow model from the joint SN+afterglow fit described by the power law $F_{\nu} \propto \nu^{-0.76\pm0.07}$.}
\label{fig:CaII}
\end{figure}

\begin{figure}[h]
\centering
\includegraphics[width=\textwidth]{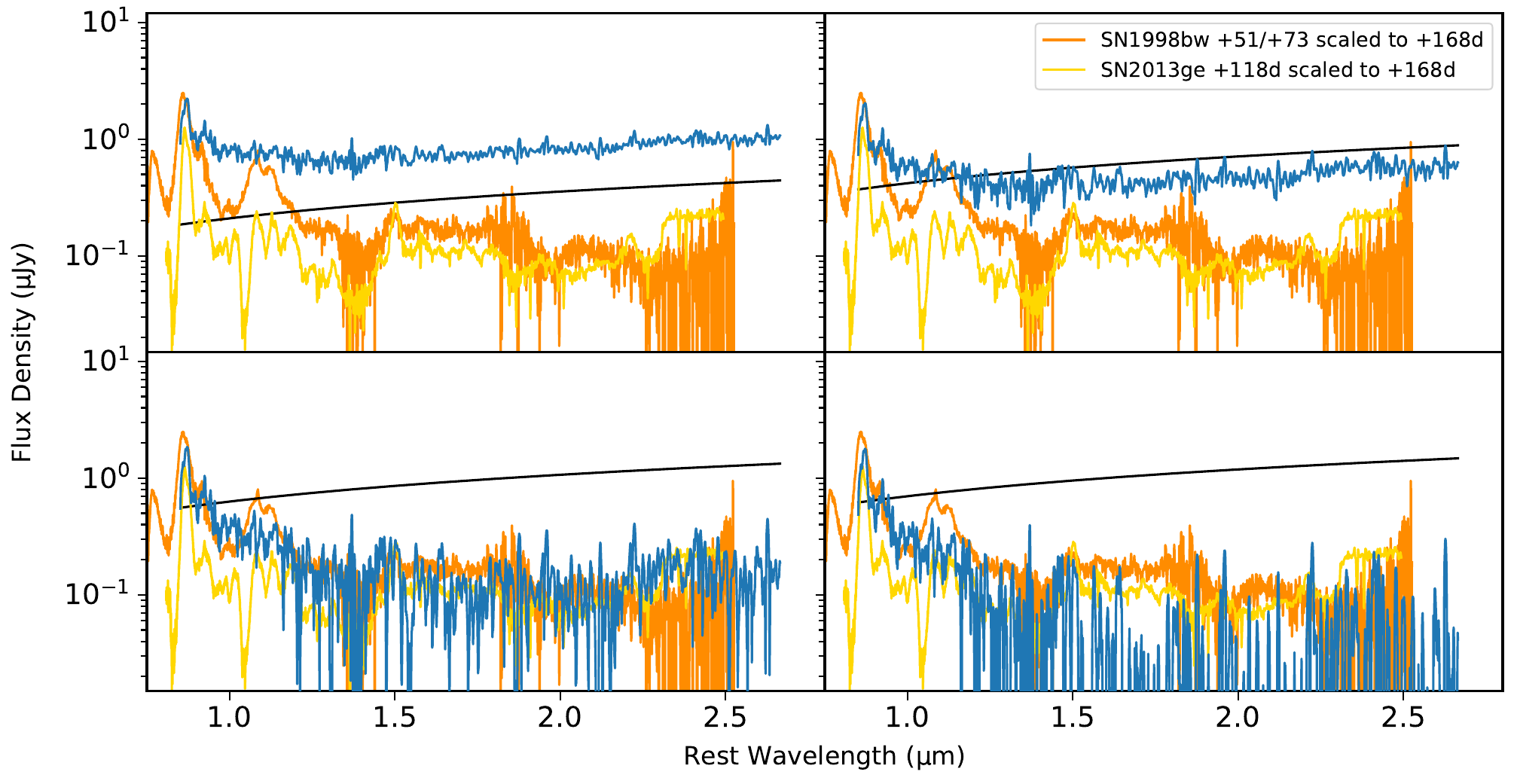}
\caption{Our extinction-corrected NIRSpec spectrum of \grb\ (blue; smoothed) with various afterglow models subtracted compared to ground-based late-time NIR spectra of SN\,1998bw (orange) and SN\,2013ge (gold) scaled to the distance of \grb\ and the same phase of our observations.  In each panel, we show our spectrum of \grb\ after subtracting our best-fit model for the afterglow (black curve) described by the power law $F_{\nu} \propto \nu^{-0.76\pm0.07}$ and scaled by factors of 0.3 (Top left), 0.6 (Top right), 0.9 (Bottom left), and 0.95 (Bottom right).}
\label{fig:speccomp1}
\end{figure}

\begin{figure}[h]
\centering
\includegraphics[width=\textwidth]{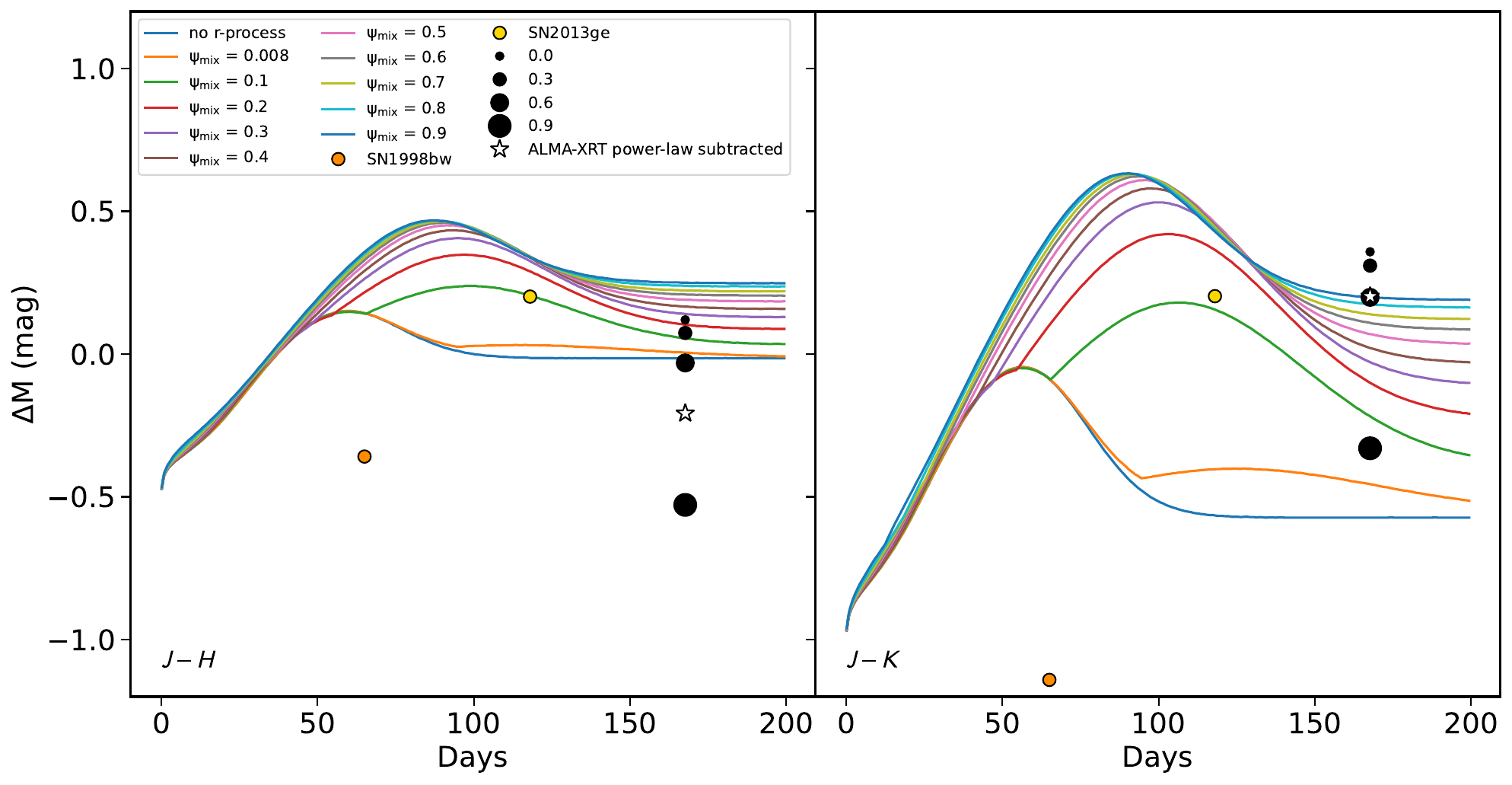}
\caption{Comparison of the $J-H$ (left) and $J-K$ (right) colors of \grb, calculated from our $+168$ day NIRSpec spectrum \textit{after} subtracting various afterglow models (star and black points), compared to the predicted color evolution of $r$-process enriched SNe from the models of \cite{BarnesMetzger2022} (lines, with $\psi$ representing the degree of mixing).  We show the colors of the SN component of our spectrum of \grb\ assuming the ALMA-XRT power law for the afterglow (star; see Extended Data Figure \ref{fig:speccomp_almaxrt}) as well as our best-fit afterglow model from fitting the red end of our spectrum scaled by factors of 0.0 (i.e.~no subtraction), 0.3, 0.6, and 0.9 (black points; see Extended Data Figure \ref{fig:speccomp1}).  We also show the colors of SN\,1998bw (orange point) and SN\,2013ge (yellow point) calculated from their spectra.}
\label{fig:rpcolors}
\end{figure}

\begin{figure}
\centering
\includegraphics[width=\textwidth]{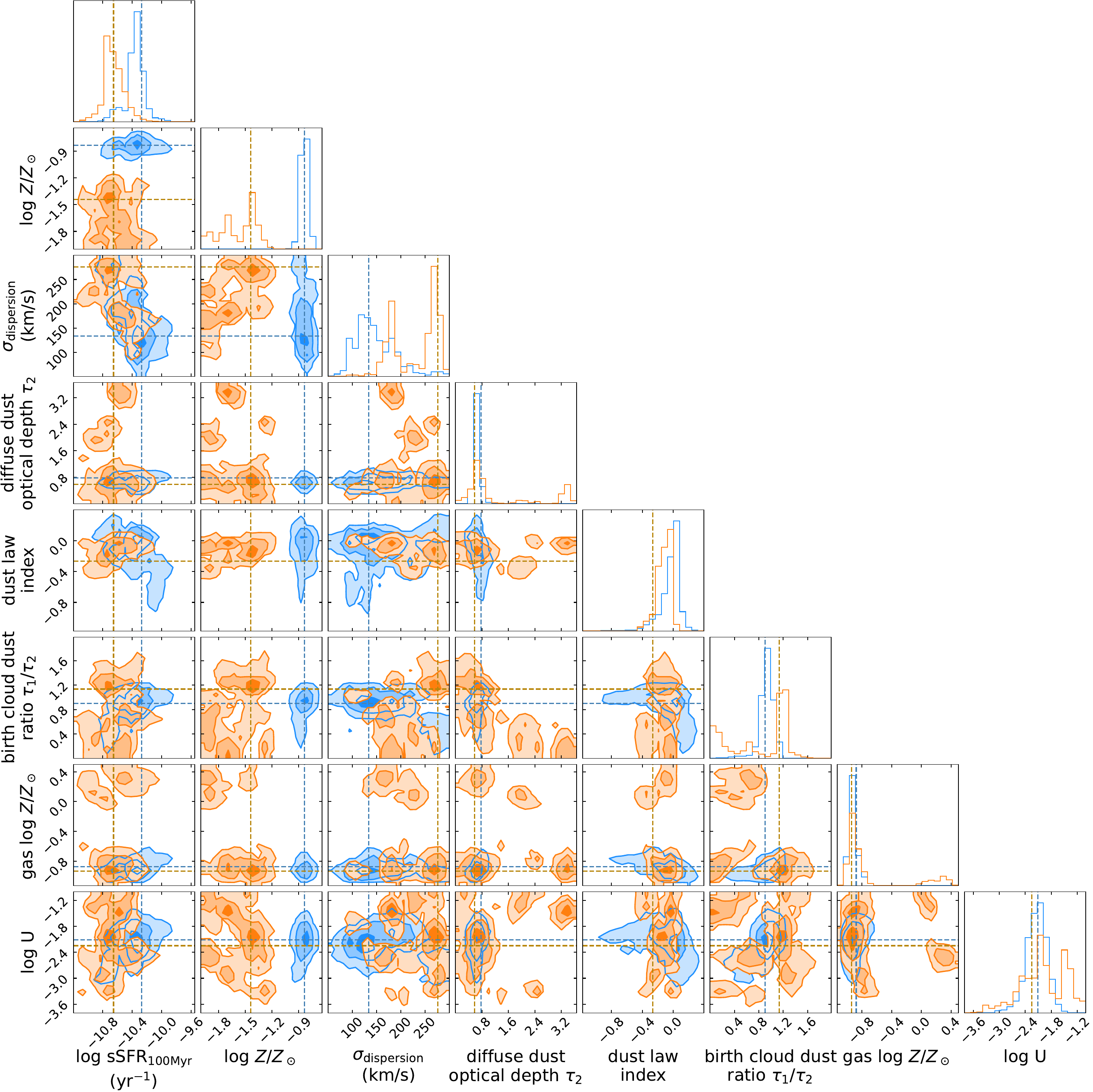}
\caption{Comparison of the \texttt{Prospector} model parameter posteriors for the galaxy spectra extracted at the location of \grb\ (orange) and the total galaxy (blue).  The gas-phase metallicities are similar, but the GRB position exhibits a lower stellar metallicity.}
\label{fig:cornerplot}
\end{figure}

\begin{table}
\caption{Best-fit Extinction Parameters and Uncertainties From Fitting the $+$13.2 day (observer frame) NIRSpec/MIRI Data} 
\label{tab:dust}
\begin{tabular}{ccccc}
\toprule
Dust Law & $\log_{10}$(Norm, Jy) & $\beta$ & $A_V$ & $R_V$ \\
\midrule
\cite{gordon2023one} & $-3.85^{+0.01}_{-0.01}$ & $0.39^{+0.01}_{-0.01}$ & $4.37^{+0.05}_{-0.05}$ & $3.07^{+0.04}_{-0.05}$\\
\cite{Fitzpatrick99} & $-3.83^{+0.01}_{-0.01}$ & $0.41^{+0.01}_{-0.01}$ & $4.63^{+0.13}_{-0.64}$ & $4.24^{+0.74}_{-0.64}$\\
\botrule
\end{tabular}
\end{table}

\begin{table}
\caption{H$_2$ Emission Line Ratios with Respect to H$_2$ 2.122 $\mu$m}
\label{tab:H2}
\begin{tabular}{cccc}
\toprule
Line ($\mu$m) & Measured Ratio & Fluorescence Model & Shock Model \\
\midrule
H$_2$ 1.233 & 1.06 $\pm$ 0.80 & 0.47 & 0.01 \\
H$_2$ 1.311 & 0.08 $\pm$ 0.26 & 0.43 & 0.00 \\
H$_2$ 1.314 & 0.92 $\pm$ 0.48 & 0.53 & 0.01 \\
H$_2$ 1.957 & 2.72 $\pm$ 1.18 & NA & NA \\
H$_2$ 2.033 & 1.33 $\pm$ 0.58 & 0.56 & 0.37 \\
H$_2$ 2.073 & 0.42 $\pm$ 0.21 & 0.25 & 0.08 \\
H$_2$ 2.223 & 1.57 $\pm$ 0.68 & 0.61 & 0.21 \\
H$_2$ 2.247 & 0.47 $\pm$ 0.21 & 0.53 & 0.08 \\
\botrule
\end{tabular}
\end{table}





\end{appendices}


\bibliography{mybib}

\end{document}